\documentclass[longauth]{aa}  
\usepackage[switch]{lineno}
\usepackage{graphicx}
\usepackage{subcaption}
\usepackage{mwe}
\usepackage{booktabs}
\usepackage{tabularx}
\usepackage{multirow}
\usepackage{natbib}
\usepackage{txfonts}
\usepackage{amsmath}
\usepackage{hyperref}
\usepackage{bbm}
\usepackage{array}
\usepackage{lscape} 
\newcolumntype{P}[1]{>{\centering\arraybackslash}p{#1}}

\usepackage{ragged2e}
\usepackage{academicons}
\usepackage{orcidlink}

\DeclareUnicodeCharacter{0308}{HERE!HERE!}
\usepackage{txfonts}

\begin{document}

   \title{{\sc{PICZL}}: Image-based photometric redshifts for AGN}


\author{William Roster\fnmsep\thanks{wroster@mpe.mpg.de}\inst{1},
M. Salvato\inst{1,2},
S. Krippendorf\inst{3,4},
A. Saxena\inst{5},
R. Shirley\inst{1},
J. Buchner\inst{1},
J. Wolf\inst{2,6},
T. Dwelly\inst{1}, \\
F. E. Bauer \inst{7,8,9,10},
J. Aird\inst{11},
C. Ricci\inst{12,13}, 
R. J. Assef \inst{12},
S.F. Anderson\inst{14},
X. Liu\inst{15,16,17},
A. Merloni\inst{1},
J. Weller\inst{1,4} \\
K. Nandra\inst{1}
}    

\titlerunning{Image-based photo-z for AGN}
\authorrunning{Roster et al.}

\institute{Max-Planck-Institut f\"ur extraterrestrische Physik, Giessenbachstr. 1, 85748 Garching, Germany
\and 
Exzellenzcluster ORIGINS, Boltzmannstr. 2, D-85748 Garching, Germany 
\and
 LMU Munich, Arnold Sommerfeld Center for Theoretical Physics, Theresienstr. 37 80333 München Germany
\and
LMU Munich, Universit\"ats-Sternwarte, Scheinerstr. 1, 81679 M\"unchen, Germany
\and
Technical University of Munich
Department of Computer Science - I26
Boltzmannstr. 3
85748 Garching b. München
Germany
\and
Max Planck Institut für Astronomie, Königstuhl 17, D-69117, Heidelberg, Germany
\and
Instituto de Astrof{\'{\i}}sica, Facultad de F{\'{i}}sica, Pontificia Universidad Cat{\'{o}}lica de Chile, Campus San Joaquín, Av. Vicuña Mackenna 4860, Macul Santiago, Chile, 7820436
\and
Centro de Astroingenier{\'{\i}}a, Facultad de F{\'{i}}sica, Pontificia Universidad Cat{\'{o}}lica de Chile, Campus San Joaquín, Av. Vicuña Mackenna 4860, Macul Santiago, Chile, 7820436
\and
Millennium Institute of Astrophysics, Nuncio Monse{\~{n}}or S{\'{o}}tero Sanz 100, Of 104, Providencia, Santiago, Chile
\and
Space Science Institute, 4750 Walnut Street, Suite 205, Boulder, Colorado 80301 
\and
Institute for Astronomy, University of Edinburgh, Royal Observatory, Edinburgh EH9 3HJ, UK
\and
Instituto de Estudios Astrof\'isicos, Universidad Diego Portales, Av. Ej\'ercito Libertador 441, Santiago 8370191, Chile
\and
Kavli Institute for Astronomy and Astrophysics, Peking University, Beijing 100871, China
\and
Department of Astronomy, University of Washington, Box 351580, Seattle, WA, 98195, USA
\and
Department of Astronomy, University of Illinois at Urbana-Champaign, Urbana, IL 61801, USA
\and
National Center for Supercomputing Applications, University of Illinois at Urbana-Champaign, Urbana, IL 61801, USA
\and
Center for Artificial Intelligence Innovation, University of Illinois at Urbana-Champaign, 1205 West Clark Street, Urbana, IL 61801, USA 
}


 
  \abstract
   {Computing reliable photometric redshifts (photo-z) for active galactic nuclei (AGN) is a challenging task, primarily due to the complex interplay between the unresolved relative emissions associated with the supermassive black hole and its host galaxy. Spectral energy distribution (SED) fitting methods, while effective for galaxies and AGN in pencil-beam surveys, face limitations in wide or all-sky surveys with fewer bands available, lacking the ability to accurately capture the AGN contribution to the SED, hindering reliable redshift estimation. This limitation is affecting the many 10s of millions of AGN detected in existing datasets, e.g., those AGN clearly singled out and identified by SRG/eROSITA.}
   {Our goal is to enhance photometric redshift performance for AGN in all-sky surveys while simultaneously simplifying the approach by avoiding the need to merge multiple data sets. Instead, we employ readily available data products from the 10th Data Release of the Imaging Legacy Survey for the Dark Energy Spectroscopic Instrument, which covers > 20,000 deg\textsuperscript{2} of extragalactic sky with deep imaging and catalog-based photometry in the \textit{grizW1-W4} bands. We fully utilize the spatial flux distribution in the vicinity of each source to produce reliable photo-z.}
   {We introduce {\sc{PICZL}}, a machine-learning algorithm leveraging an ensemble of convolutional neural networks. Utilizing a cross-channel approach, the algorithm integrates distinct SED features from images with those obtained from catalog-level data. Full probability distributions are achieved via the integration of Gaussian mixture models.}
   {On a validation sample of 8098 AGN, {\sc{PICZL}} achieves an accuracy $\sigma_{\mathrm{NMAD}}$ of 4.5\% with an outlier fraction $\eta$ of 5.6\%. These results significantly outperform previous attempts to compute accurate photo-z for AGN using machine learning. We highlight that the model's performance depends on many variables, predominantly the depth of the data and associated photometric error. A thorough evaluation of these dependencies is presented in the paper.}  
  {Our streamlined methodology maintains consistent performance across the entire survey area, when accounting for differing data quality. The same approach can be adopted for future deep photometric surveys such as LSST and Euclid, showcasing its potential for wide-scale realization. With this paper, we release updated photo-z (including errors) for the XMM-SERVS W-CDF-S, ELAIS-S1 and LSS fields.}
  

   \maketitle
%

\section{Introduction}

In recent decades, our understanding of Active Galactic Nuclei (AGN) and their role in galaxy and cosmic evolution has significantly advanced. These luminous celestial powerhouses are thought to be fueled by the accretion of matter onto supermassive black holes (SMBHs) located at the centers of galaxies, exerting intense energetic radiation across the entire electromagnetic spectrum, ranging from radio to $\gamma$-rays \citep{Padovani2017}. The close correlation observed between the mass of the central SMBH, whether active or inactive and the properties of its host galaxy's bulge — such as the galaxy's mass and velocity dispersion \citep[e.g.,][]{Gebhardt2000, Ferrarese2000} — suggests a co-evolutionary relationship between galaxies and their central engines \citep{Kormendy2013, Heckman2014}. Ongoing research focuses on understanding scaling relations, the evolution of SMBHs within galaxies, and the interconnected rates of star formation (SFR) and black hole accretion (BHAR) over cosmic time \citep[e.g.,][]{Madau2014}. To further explore and address these unresolved topics requires diverse AGN samples with reliable redshifts to determine BH demographics and constrain models of galaxy evolution. For all these studies, redshift is an indispensable quantity, with spectroscopic redshifts (spec-z) remaining the preferred estimates for determining precise cosmic distances \citep{hoyle2016}. However, while multi-objects spectrographs, such as the Sloan Digital Sky Survey \citep[SDSS-V;][]{York2000, Kollmeier2019SDSS},
the Dark Energy Spectroscopic Instrument \citep[DESI;][]{desicollaboration2016desi}, the Subaru Prime Focus Spectrograph \citep[PFS;][]{Tamura2016} or the 4-metre Multi-Object Spectroscopic Telescope \citep[4MOST;][]{Jong2019}, are set to provide a drastic rise in the number of observed sources over the next several years, we are currently in the situation in which millions of AGN have been detected all-sky by various surveys (e.g., by the Wide-field Infrared Survey Explorer mission \citep[WISE;][]{Wright2010}, and the extended Roentgen Survey with an Imaging Telescope Array \citep[eROSITA;][]{merloni2012erosita,Predehl2021}, with only the brightest sources having been observed spectroscopically \citep{Dahlen2013}. The growing disparity between photometric and spectroscopic observations will only widen with upcoming surveys such as the Legacy Survey of Space and Time \citep[LSST;][]{Ivezi2019} and Euclid \citep{euclidcollaboration2024euclid}, covering unprecedented areas and depths \citep{NewmanGruen2022}. Thus for the bulk of AGN, we must make use of multiband photometry and rely on photometric redshifts (photo-z).

First implemented by \cite{Baum1957} for inactive galaxies, these low-precision redshift estimates utilize photometric observations to effectively obtain a sparsely sampled spectral energy distribution (SED), trading precision for scalability. They encompass an array of techniques assuming color-redshift evolution \citep{Connolly1995,Steidel1996,Illingworth1999,Bell2004}, including template-based approaches \citep[e.g.,][]{Bolzonella2000,Ilbert2006,Salvato2008,Beck2017}, where redshifted models built on theoretical or empirical SEDs are fitted to observed multi-band photometry. Although a limited number of available bands can introduce uncertainties \citep[see review by][]{salvato2018}, photo-z methods offer an efficient way to estimate distances for all sources in an imaging survey, yielding highly accurate estimates with as few as three bands for passive galaxies \citep{Benitez2000,Abdalla2011, Arnouts,Brescia2014,Desprez2020}. By contrast, reliable photo-z for AGN have historically required highly homogenized photometry across >20 filters, which was only achievable in pencil-beam surveys \citep{Salvato2011}. As such, this level of detail continues to be unfeasible for wide-area surveys. However, with the 10th data release of the DESI Legacy Imaging Surveys \citep[LS10,][]{Dey2019}, we now have a broad-sky survey that, while lacking NIR coverage, includes a few optical bands supplemented by mid-IR WISE data. This allows us to explore the possibility of generating reliable photo-z for AGN over the full sky, despite having fewer filters compared to the densely sampled pencil-beam surveys.

SED fitting applied to a broad population of AGN remains particularly challenging due to the uncertainty and difficulty of disentangling the relative contributions of the nucleus and respective host to a given band \citep[e.g.,][]{Luo2010, Salvato2011, Brescia2019}. Since the accretion properties of SMBHs, often characterized as the bolometric luminosity divided by the Eddington limit, or the Eddington ratio, significantly influence the SED of AGN, the intense power-law continuum radiation can either partly (host-dominated) or entirely (quasar-dominated), outshine the respective host, hiding key spectral features that lead to redshift degeneracies \citep{Pierce2010, Povic2012, Bettoni2015}. Consequently, selecting a limited number of templates can be insufficient for correct redshift determination, while increasing the number of templates raises the degeneracy \citep[see discussion in][]{Salvato2011,Ananna2017}. In this regime of accounting for AGN contributions to galaxy photo-z, one potential approach involves modeling objects as a combination of quasar and galaxy templates \citep[eg.,][]{Cardamone2010}, performed with EAZY \citep{Brammer2008}. In addition, surveys typically estimate fluxes with models that do not account for a mixed contribution from AGN and host galaxy. Ultimately, AGN are also intrinsically variable sources on the timescales explored by the previously mentioned surveys leading to incongruent photometry acquired across different epochs.

In contrast to template-fitting methods, more recent approaches have shifted towards the use of empirical Machine Learning (ML) models, performing regression or classification, to tackle photo-z applied predominantly to inactive galaxies \citep{Collister2004,Laurino2011,Zhang2013,hoyle2016,DIsanto2018,Brescia2019,Eriksen2020,Li2021}. Provided with a very large and complete spec-z sample, ML architectures manipulate photometric input features to minimize the divergence between spectroscopic and ML-derived redshifts.
Over the years, a plethora of ML architectures, including decision trees \citep{Breiman2001, Carliles2010,Li2022}, Gaussian processes \citep{Almosallam2016} and K-nearest neighbours \citep{Zhang2013, Luken2019} have been employed, yielding accurate point predictions and, more interestingly, full probability density functions (PDFs) \citep{CarrascoKind2013, Cavuoti2016, rau2015accurate, Sadeh2016}. The latter grants access to the prediction uncertainty, as otherwise naturally provided by template-fitting approaches, relevant for studies dealing with, e.g. luminosity functions \citep{Aird2010,Buchner2015,Georgakakis2015}. However, the limited availability of a sizable training sample of AGN has resulted in only a few attempts to compute photo-z for mostly nucleus-dominated objects with ML-based methods \citep{Mountrichas2017,Fotopoulou2018,Ruiz2018,Meshcheryakov2018,Brescia2019,nishizawa2020photometric}.

More recently, the conventional approach of manually selecting photometric features for ML has been replaced by bright, well-resolved galaxies at low redshift \citep{hoyle2016,Pasquet2018,campagne2020adversarial,Hayat2021}. In this regime, integrating galaxy images into deep neural networks inherently captures essential details like flux, morphology, and other features that would typically be extracted from catalogs based on predefined assumptions, leading to a more comprehensive redshift estimation process. This approach is particularly advantageous for addressing current limitations faced by photo-z methods for AGN, as it leverages model-independent fluxes and redshift indicative features, including surface brightness profiles \citep{Stabenau2008,Jones2017,Gomes2017,Zhou2021,Zhou2023}. Unlike creating a single SED from total flux measurements, projects employing images with independent pixel-by-pixel SEDs at identical redshift have demonstrated increased photo-z constraining power, alleviating previous empirical approaches by decreasing the fraction of outliers \citep{Henghes2021,Schuldt2021,Lin2022,Dey2022,NewmanGruen2022}.

Here, we introduce {\sc{PICZL}} (Photometrically Inferred CNN redshift(Z) Likelihoods), an enhanced approach to photo-$z$ estimation that builds upon \cite[{\sc{CircleZ}} by][]{saxena2024}. While the authors demonstrated that redshift degeneracies encountered for AGN, typical in cases of limited photometry, can be broken by integrating aperture photometry alongside traditional total/model fluxes and colors, {\sc{PICZL}} instead computes photo-z PDFs for AGN directly from homogenized flux band cutouts by leveraging the more detailed spatial light profile. All inputs are obtained utilizing LS10 exclusively. Similar to \cite{saxena2024}, {\sc{PICZL}} can produce reliable photo-z PDFs for all Legacy-detected sources associated with an AGN. However, the model can, in principle, be applied to other extragalactic sources (e.g, inactive galaxies, Götzenberger et al. in prep.) granted that a dedicated training sample is used.

We employ an ensemble of the same ML algorithm, notably convolutional neural networks (CNNs), known for their proficiency in learning intricate patterns, as outlined by \citep{Lecun1998}. Specifically designed for image analysis, CNNs excel at identifying and extracting relevant predictive features directly from images, thereby reducing computational overhead compared to fully connected architectures. Harnessing this more extensive pool of information, these models surpass alternative models based on condensed feature-based input sets.

The paper is structured as follows: Sect. \ref{sec2} introduces the AGN training sample down-selection. Sect. \ref{sec3} focuses on the photometric data products available within LS10. Sect. \ref{sec4} details the photometric data preprocessing, followed by Sect. \ref{sec5}, which outlines the model pipeline. Sect. \ref{sec6} presents and quantifies the redshift results, while Sect. \ref{sec9} evaluates the photo-z released for the XMM-SERVS \citep{Chen2018,Ni2021} fields. Sect. \ref{sec10} outlines current limitations and discusses how we can achieve further improvements. Sect. \ref{sec11} explores implications for future surveys, concluding with a summary.

In this paper, unless stated differently, we express magnitudes in the AB system and adopt a $\Lambda$CDM cosmology with $H_{0} = 69.8$ km s\textsuperscript{-1} Mpc\textsuperscript{-1}, $\Omega_{\text{m}} = 0.28 $ and $\Lambda = 0.72$.


\section{AGN training sample selection} 
\label{sec2}

In X-ray surveys, the identification of AGN has two distinct advantages - i) the reduced impact of moderate obscuration and ii) the lack of host dilution. Due to the inherent brightness of accreting SMBHs compared to their host galaxies, this results in a significantly higher nuclei-to-host emission ratio compared to observations in some of the neighbouring wavelength windows, such as UV-optical-NIR \citep{Padovani2017}. This naturally leads to a larger diversity of AGN observed by an X-ray telescope. That being said, surveys in the more accessible optical and NIR regime can increase the likelihood of detecting higher-$z$, and in the case of MIR more heavily obscured AGN, compared to the soft X-ray bands.

ML approaches for photo-$z$ estimation in large surveys \citep[e.g.,][]{Fotopoulou2018,Duncan2022} typically classify objects into three broad categories: galaxies, quasars (QSOs), and stars, before computing photo-z. However, this classification is usually based on the optical properties and hence fails for obscured and/or lower-luminosity AGN. Since our goal is to improve on the quality of photo-z estimates for X-ray detected extragalactic sources, including type 2 AGN and low-redshift Seyfert 1 galaxies, generally, our training sample has to replicate this diversity. We achieve this by combining AGN selected across multiple wavelength bands.

As a starting point, we include the same X-ray samples used in \citet{saxena2024}, namely the latest version of the XMM catalog, 4XMM, which spans 19 years of observations made with \textit{XMM-Newton} \citep{Webb2020} and data from the eROSITA CalPV-phase Final Equatorial-Depth Survey \citep[eFEDS;][]{Brunner2022}, as they provide a reasonably representative and complete set of diverse AGN spanning 5 dex in X-ray flux out to redshift $z \lesssim 4$. However, with just these, some portions of AGN parameter space remain imbalanced, such as highly luminious and/or high-z AGN. Thus, we expand the dataset by adding bright, optical and MIR, selections. We describe each of these samples in more detail below.

This approach enhances the completeness of our training sample, which is essential to mitigate covariate shift, i.e., the shift in parameter space between the training and validation samples, so that the model generalizes effectively to new data \citep{Norris2019}. Subsequently, a non-representative training sample may lead to systematically biased outcomes \citep{NewmanGruen2022}. Accordingly, algorithms will be strongly weighted towards the most densely populated regions of the training space \citep{Duncan2022}.

\begin{figure}[h!]
\centering
\includegraphics[width=9cm]{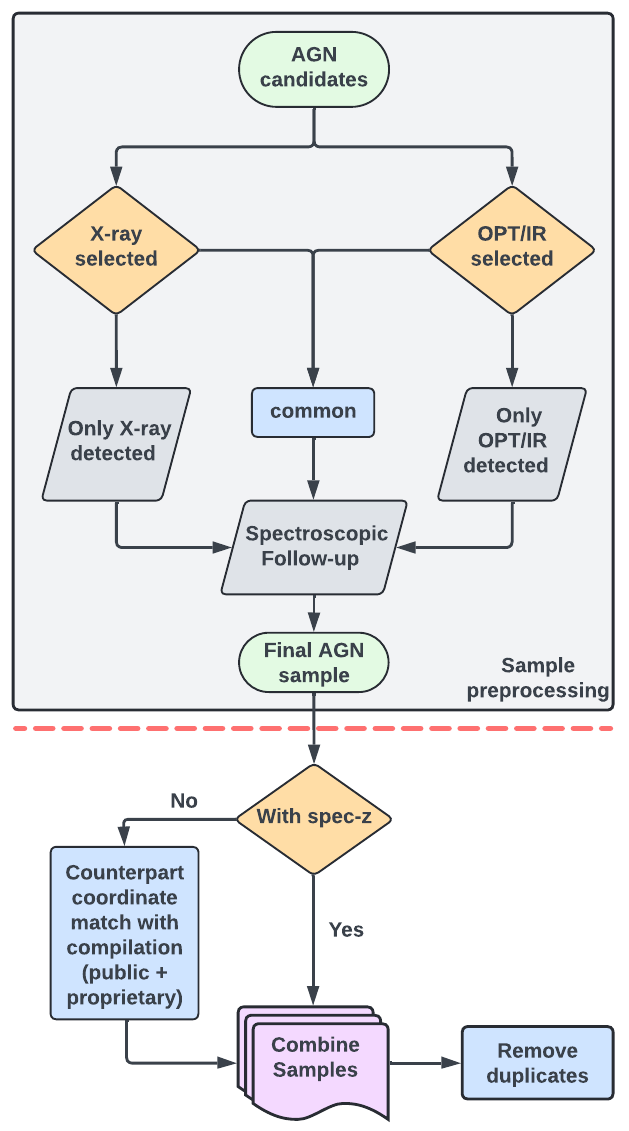}
\caption{Flowchart depicting the training sample down-selection pipeline. This includes the sample preprocessing (grey box) and the sample refinement, including redshift extension and duplicate removal below the dashed red line.}
\label{fig:2.8}
\end{figure}

\subsection{Beyond eFEDS and 4XMM}
\label{noxsamples}

We can enhance the redshift distribution within our sample, particularly towards high ($z\geq 3$) redshift, in this otherwise underrepresented parameter space due to observational selection effects. We recognize the subsequent incorporation of unavoidable selection biases in each survey while restricting the inclusion of sources at low redshift to a minimum. While the balance between dataset quality and size is critical, deep learning algorithms that operate on pixel-level inputs tend to perform optimally only when training datasets contain $\geq$ 400\ 000 galaxy images \citep{Schuldt2021,Dey2022, NewmanGruen2022}. Since our method would benefit from a larger sample (see Table \ref{Tab_2.1}), we chose not to apply stringent quality criteria by only considering high-quality data, significantly reducing the number of sources available training. Such a reduction would also prevent the model from learning to handle lower-quality data, limiting its application to only high-quality validation data. By not making an initial down-selection, we retain the flexibility to apply quality cuts to future blind samples by using LS10 flags later.

\begin{table}
\centering
\caption{Overview of the relative fractions of source catalogs used in compiling our AGN training sample.}
\begin{tabular}{P{1.6cm}|P{2.2cm}|P{3.4cm}}
\hline
&&\\
Selection & Source catalog & Unique LS10 \& spec-z \\
&&\\ 
\hline
\hline
&& \\
X-ray & eFEDS &  9931 (24.5\%) \\
 & 4XMM &   7022 (17.3\%) \\
  & SDSS-V &  401 (1\%)\\
&&\\ 
Optical & 2QZ &   20500 (50.6\%)\\
 & DESI EDR &   349 (0.9\%)\\
 & QUBRICS &  28 (0.1\%)\\
&&\\ 
Composite  & DR16Q &   1814 (4.5\%) \\
  & High-Z &  444 (1.1\%) \\
&&\\ 
\hline
\end{tabular}
\label{Tab_2.1}
\end{table}

\subsubsection{Samples from optical selection}

We include the 2dF QSO Redshift Survey \citep[2QZ,][]{Croom2004} with $\sim$ 23k color selected QSOs in the magnitude range 18.25 $\leq$ b\textsubscript{J} $\leq$ 20.85 at redshifts lower than $ z \sim 3$ and the QUasars as BRIght beacons for Cosmology in the Southern hemisphere survey \citep[QUBRICS,][]{Boutsia2020} with 224 bright (i $\leq$ 18) QSOs at redshifts of $z \geq 2.5$. 

\subsubsection{Samples from optical follow-up of X-ray sources}

Additionally, we incorporate the SDSS-IV \citep{Blanton2017} quasar catalog from Data Release 16 \citep[DR16Q,][]{Lyke2020} with $\sim$ 150k quasars collected from various subprogrammes including optical/IR selection down to g $\leq$ 22 in the LS10 footprint after subselecting high quality spec-$z$, as well as follow-up of X-ray sources from ROSAT \citep{voges1999rosat,Boller2016, salvato2018} and XMM \citep[e.g.][]{LaMassa2019}. As successor science programme, we also consider the Black Hole Mapper (BHM) SPectroscopic IDentfication of ERosita Sources (SPIDERS, Anderson et al., in prep, Aydar et al., in prep) from SDSS-V (Kollmeier et al., in prep) Data Release 18 \citep{dwelly2017,Coffey2019,Comparat2020,Almeida2023}.

\subsubsection{Samples of high-z sources}

Given the strong imbalance above $z \sim 3.5$, we also include 400 optically/IR selected quasars at redshifts 4.8 $\leq z \leq 6.6$ down to g $\leq$ 24 from the high-redshift quasar survey in the DESI Early Data Release \citep[EDR,][]{Yang2023} and a compilation of high-z quasars at $z \geq 5.3$ published in literature \citep{fan2022quasars}.

\subsection{Spectroscopic cross-referencing}

\begin{figure}[h!]
\centering
\includegraphics[width=9cm]{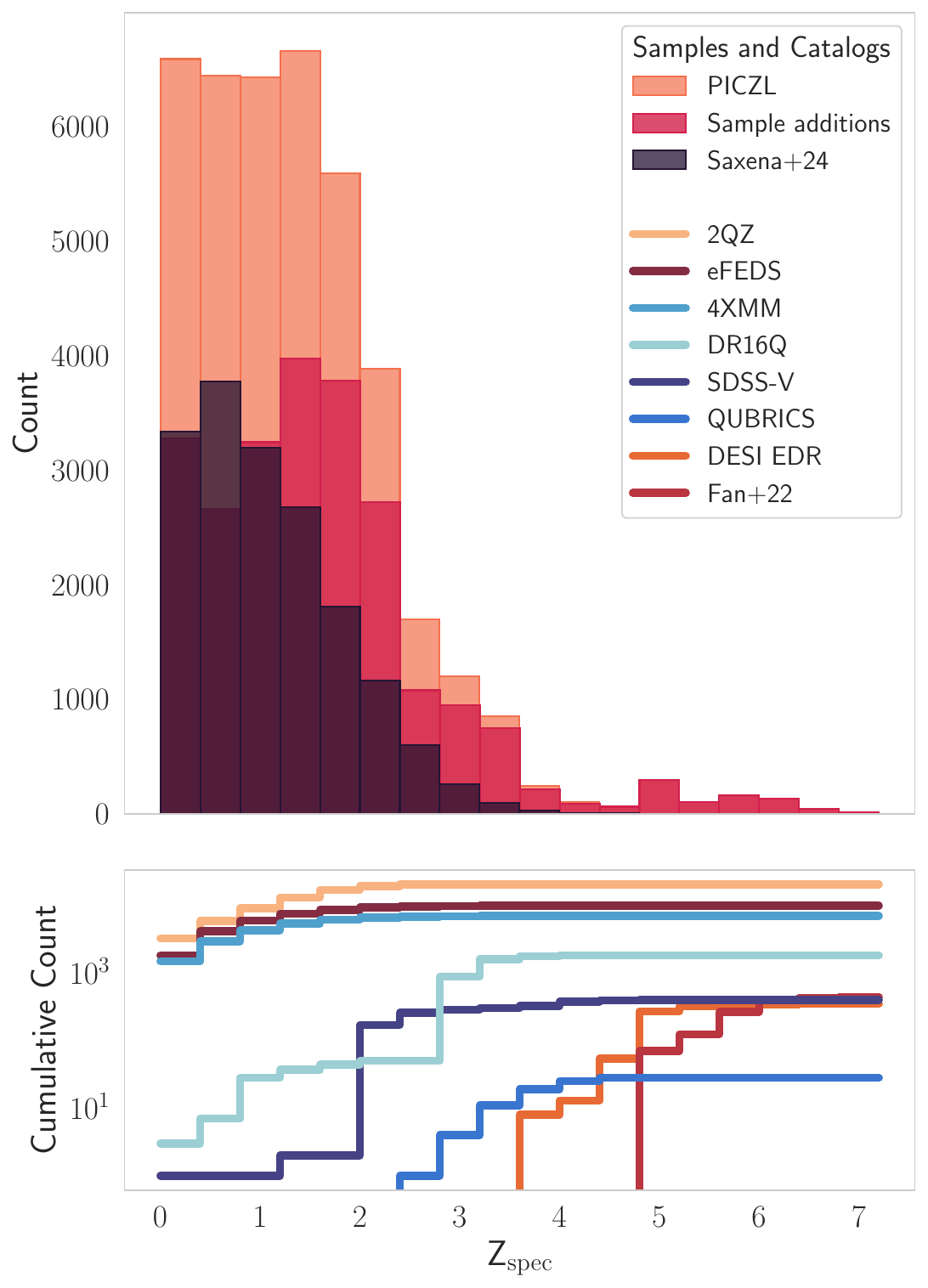}
\caption{Binned redshift histogram (top panel) and cumulative distribution (bottom panel) of the sources utilized in the {\sc{PICZL}} AGN sample. Note that these samples are not necessarily rank ordered by importance but for improved readability.}
\label{fig:2.3}
\end{figure}

The parent sample of AGN is annotated with spec-$z$, where available. According to Figure \ref{fig:2.8}, we also consider sources, including those from eFEDS and 4XMM, with spatial counterparts from a compilation of public redshifts \citep{kluge2024}. The procedure by which we match optical counterparts in our combined sample to a compilation of quality criteria down-selected spec-$z$, is outlined in Sect. 3.1 of \cite{saxena2024}. Due to overlaps between surveys, we remove duplicates when combining samples. The final sample of sources with spec-$z$ comprises 40$\ $489 objects, with a breakdown in Table \ref{Tab_2.1}. Correspondingly, the (cumulative) histograms illustrating the n($z$) distributions that collectively constitute the {\sc{PICZL}} sample are presented in Figure \ref{fig:2.3}.

\section{The survey}
\label{sec3}

To streamline and simplify our methodology, we have 
chosen to employ data from LS10 exclusively to mitigate potential complications arising from the heterogeneity of multiple datasets. Crucially, the survey area now extends over 20$\ $000 deg\textsuperscript{2}
of optical $griz$ and WISE $W1-W4$ forced photometry, by incorporating the following datasets:

\begin{itemize}
    \item DECam Legacy Survey observations \citep[DECaLS,][]{Flaugher_2015,Dey2019}, including data from the Dark Energy Survey \citep[DES,][]{Abbott2016}, which covers a 5000 deg\textsuperscript{2} contiguous area in the South Galactic Cap. In the DES area, the depth reached is higher than elsewhere in the footprint.
    \item DECam observations from a range of non-DECaLS surveys, including the full six years of the Dark Energy Survey, publicly released DECam imaging (NOIRLab Archive) from other projects, including the DECam Local Volume Exploration Survey \citep[DELVE,][]{DrlicaWagner2021} and the DECam eROSITA survey (DeROSITAs, PI: A. Zenteno, Zenteno et al. in prep).
\end{itemize}
In the north ($\delta$ > 32.375  deg), LS10 uses the Beijing-Arizona Sky Survey \citep[BASS,][]{Zou17} for $g$- and $r$-band coverage, and the Mayall $z$-band Legacy Survey \citep[MzLS,][]{Silva2016} for $z$-band coverage \citep{kluge2024}.

\begin{figure*}
    \centering
    \includegraphics[width=18cm]{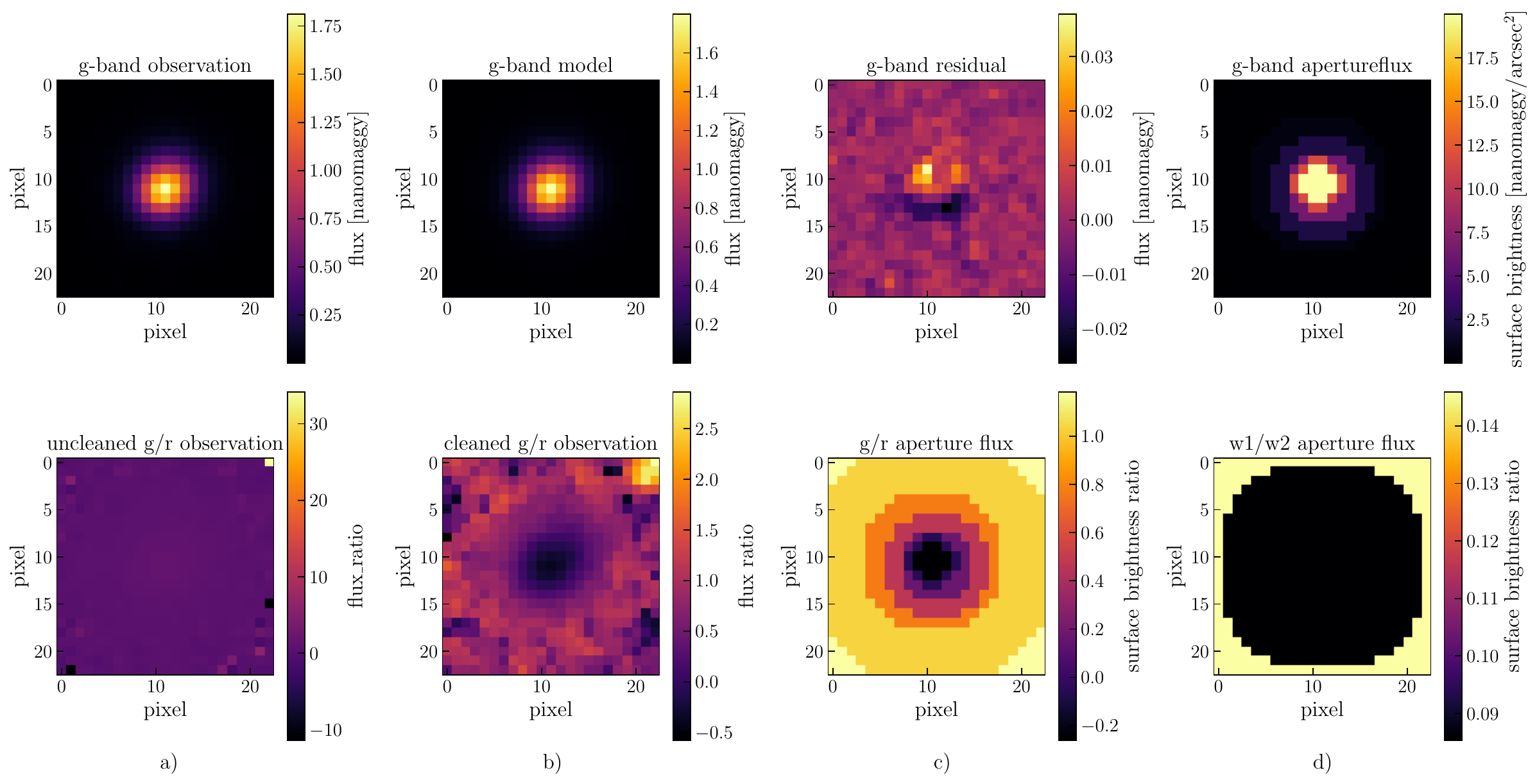}
    \caption{Grid of subplots showing various model inputs for {\sc{PICZL}}. In the upper row: the LS10 g-band image (a), along with its 2-D model flux (b), residual (c), and aperture flux map (d). In the bottom row: the original g-r color is shown in (a). The presence of a saturated pixel in the top right corner is visible, indicating the need for pre-processing. The result of the preprocessing is shown in b).  The bottom panels c) and d) show the g/r-band and w1/w2-band aperture flux maps, respectively.}
    \label{fig:collage}
\end{figure*}

\subsection{Photometric data}

LS10 offers registered, background-subtracted, and photometrically calibrated point spread function (PSF)-forced photometry, including corresponding errors. To extend their wavelength coverage, DR10 catalogs incorporate mid-infrared (mid-IR) forced photometry at wavelengths of 3.4, 4.6, 12, and 22 $\mu$m (referred to as W1, W2, W3, and W4, respectively) for all optically detected sources in the LS10 via the Near-Earth Object Wide-field Infrared Survey Explorer (NEOWISE) \citep{Mainzer2011,Lang2014,Meisner2017}. Sources are modeled simultaneously across all optical bands, ensuring consistency in shape and size measurements by fitting a set of light profiles, even for spatially extended sources. Consequently, alongside reliable total and multi-aperture (8 annuli $\leq$ 7 arcseconds, five annuli $\leq$ 11 arcseconds for the optical and mid-infrared bands, respectively) flux measurements, the LS10 catalog offers seeing-convolved PSF, de Vaucouleurs, exponential disk, or composite de Vaucouleurs + exponential disk models obtained with the Tractor algorithm \citep{Lang2016TheTP}. 
Additionally, providing fluxes rather than magnitudes, enables considering sources with very low signal-to-noise ratios without introducing biases at faint levels. This characteristic also facilitates flux stacking at the catalog level, enhancing the overall versatility and utility of the classification and fitting process within LS10 \footnote{\url{https://www.legacysurvey.org/dr10/catalogs/}}.

\subsection{Imaging data}
\label{imgdat}

In addition to catalog data, LS10 provides a rich set of imaging products. These include observations, flux model images, and residual maps for all available bands. For instance, the top panels (a, b, c) of Figure \ref{fig:collage} display all image products for a  $g$-band observation, respectively.


\section{Preprocessing}
\label{sec4}

Here we detail the preprocessing steps taken to prepare our dataset, ensuring that it is clean, normalized, and structured appropriately.

\subsection{Image preprocessing}
\label{IP}

Building on the approach of \citet[][]{saxena2024}, which demonstrated significant improvements by shifting from total to aperture flux utilizing information on the 2D light distribution, we aim to further refine the spatial characterization of sources. This is achieved by incorporating pixel-level flux resolution through imaging as base input. Images in individual bands or in combination (i.e., colors) reflect the surface brightness, angular size, and sub-component structures of the sources, indirectly providing redshift information \citep{Stabenau2008}. To obtain reliable photo-z directly from images, we utilize flux-calibrated optical cutouts across as many filters as available.

With an average seeing FWHM of 1.3 arcseconds under nominal conditions, LS10 provides a pixel resolution of 0.262 arcseconds per pixel for the optical bands, reaching depths between 23 and 24.7 AB, depending on the specific band and region of the sky (see \citet{Dey2019} and Figure 1 in \citet{saxena2024}). To enhance computational efficiency and mitigate contamination from nearby sources, we restrict our cutout dimensions to 23$\times$23 pixel, centered on the AGN coordinates in the four $griz$ LS10 bands. Our cutouts correspond to a field of view (FOV) of approximately 6 arcseconds $\times$ 6 arcseconds. We base our choice of FOV on the angular size-redshift relation by computing the angular diameter distance $d_{\Lambda}$ via:

\begin{equation}
\frac{c}{H_{0}} \frac{1}{(1+z)} \int_{0}^{z} \frac{\textrm{d}z'}{\sqrt{\Omega_{\text{M}}(1+z')^{3}+\Omega_{\Lambda}}} \ .
\label{eq1}
\end{equation}

\begin{figure}[t!]
\centering
\includegraphics[width=9cm]{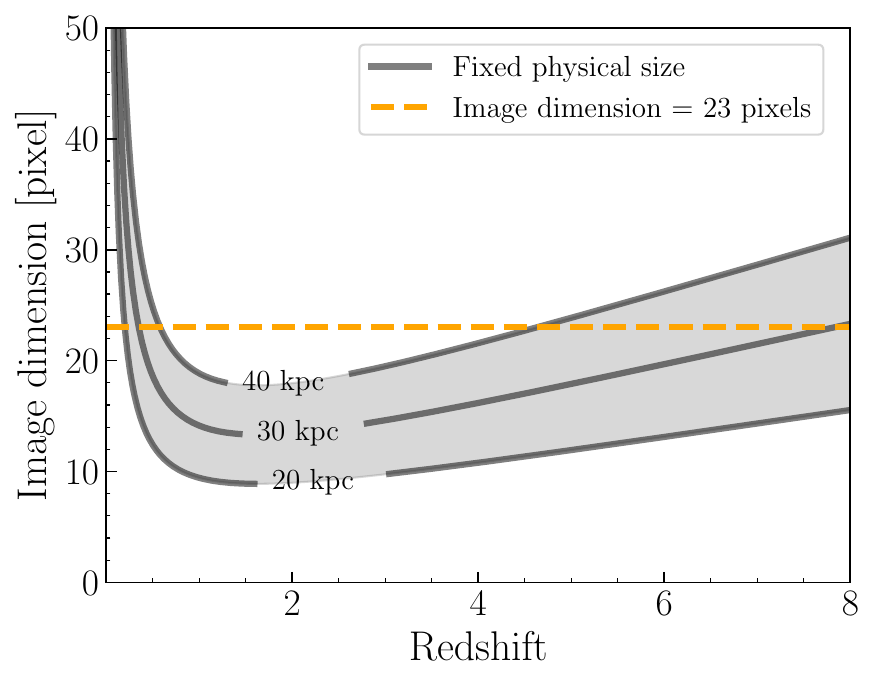}
\caption{Angular size of sources with a fixed physical size of 20--40 kpc, as a function of redshift. The orange dashed line depicts a fixed image dimension of 23 pixels, assuming the LS10 spatial resolution of 0.262 arc seconds per pixel, which suffices to cover objects of 30 kpc diameters in size out to redshifts of $\sim 0.3 \leq z \leq 7.8$.}
\label{fig:2.4}
\end{figure}

\noindent Equation \ref{eq1} and Figure \ref{fig:2.4} elucidate the connection between an object's physical size, its angular size, and redshift. Notably, we can effectively map galaxies with a diameter of 30 kpc — representative of main sequence galaxies \citep{Wuyts2011} — within the confines of a 23$\times$23 pixel cutout, covering the range of $0.5 \leq z \leq 7.7$.

\begin{figure}[t!]
\centering
\includegraphics[width=4.3cm]{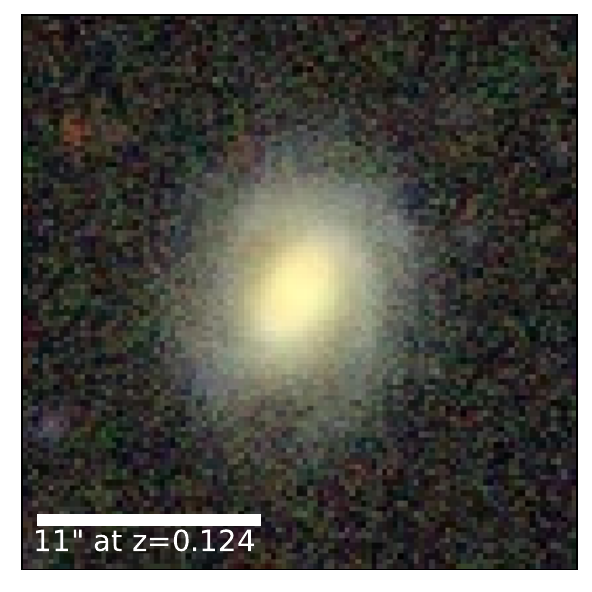}
\includegraphics[width=4.3cm]{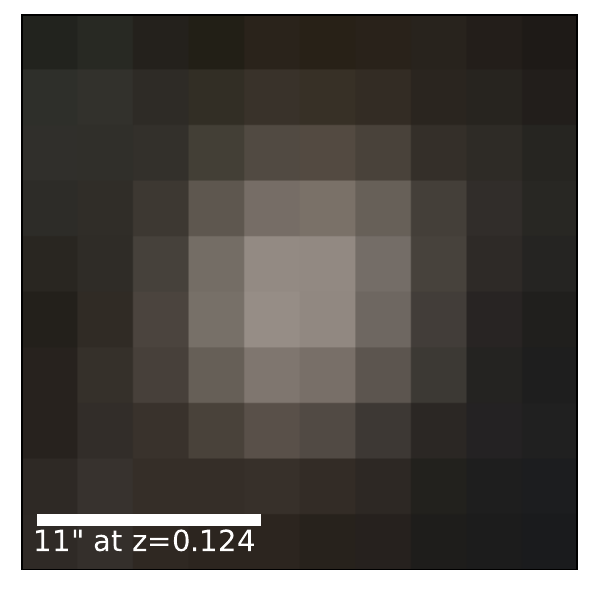}
\caption{Example AGN in our sample, as seen by LS10, in the optical RGB image (left) and in the W1 image from NEOWISE7 (right). The spatial resolution is 0.262 arcseconds and 2.75 arcseconds per pixel, respectively. The size of the cutouts is 27.5 arc seconds $\times $ 27.5 arc seconds.}

\label{fig:2.5}
\end{figure}

Given that the FWHM of the  W1, W2, and W3 images is 6 arcseconds, and of 12 arcseconds for W4, WISE band cutouts do not provide meaningful spatial information at this scale (see Figure \ref{fig:2.5}). Therefore, we have opted to use images solely from the optical bands. Problematic sources, exhibiting signs of defects in various ways are flagged in the Legacy Survey by specific bitmasks \citep{Dey2019}.

We acknowledge that LS10 images exhibit varying quality due to differences in seeing conditions during observations taken over many years, which impact in particular the measurements of color within source apertures. Despite this, we rely on the model's ability to adapt to these intrinsic variations given the size and diversity of our training sample, as the sources withheld from training represent a shuffled subsample of the main dataset, ensuring robust evaluation. We have verified that the distribution in seeing quality (expressed as the weighted average PSF FWHM of the images) for the training and validation are comparable (refer to Figure \ref{fig:G.1}). However, preliminary tests indicate that adding PSF size and PSF depth of the observations—both available in LS10—as additional features enhances model performance (Götzenberger et al., in prep.). While it is not unreasonable to expect the model to implicitly infer the PSF or a related abstract representation thereof from the images themselves, these features will be included by default in future PICZL versions. With ongoing developments, PSF cutouts are expected to become more accessible for integration into the image stack (see Table \ref{Tab_2.2}). In the longer term, we anticipate that upcoming surveys like LSST, with their improved consistency in image quality, will further reduce these limitations and boost the precision of pixel-based analyses such as ours.

\begin{table*}
\centering
\caption{Overview of the various feature formats and dimensions utilized in {\sc{PICZL}}'s multi-channel approach.}

\begin{tabular}{llll}

\hline
Data type & flux images & color images & numerical \\
Shape & (23,23,32) & (23,23,24) & (52,) \\
\hline
 & &  \\
 
Feature & observational $griz$ (4) & observational $griz$ (6)&   source type (5)\\

&aperture flux $griz$  (4) & aperture flux $griz$ (6) &  dchisq (5)\\

&aperture flux $W1-W4$  (4) & aperture flux $W1-W4$ (6) & total flux (4)\\

&residual $griz$  (4)&  model $griz$  (6) &  observed source flux fraction (8)\\

&residual aperture flux $W1-W4$  (4)& & S/N (8)\\

&aperture flux $griz$ ivar. (4)& & total flux colors (22) \\

&aperture flux $W1-W4$ ivar. (4) & &  \\

&model $griz$ (4)  & &  \\

 & &  \\

\hline
   
\end{tabular}
\tablefoot{Each column represents a distinct processing channel, detailing its input data type and the respective dimensions. All flux-related inputs have been corrected for reddening effects.}
\label{Tab_2.2}
\end{table*}

\subsection{Color images}
\label{col_im}

It proves advantageous to provide the network with color images (ratio between images from different bands) as an additional input. This approach avoids the necessity for the model to learn the significance of colors solely from the flux images, which is inherently a more difficult task. Likewise, rather than processing numerical features separately and merging them with the information extracted from the image cube at a later stage, we find it beneficial to integrate them directly at the pixel level. As a result, whenever possible, we transform catalog features into 2D arrays to align them with the original images in the same thread, enabling smoother integration and more coherent analysis \citep[see e.g.][]{Hayat2021}. We improve the depth of our data cube by converting catalog-based quantities, e.g., flux measurements from apertures across different bands, into synthetic images, with a respective image size depending on the cutout size (see Figure \ref{fig:collage}). We expand this approach by generating images for all viable color combinations of aperture fluxes, constrained to those with matching aperture sizes. Additionally, we produce color images for flux cutouts where pixel resolutions are consistent (see lower panels b) and c) of Figure \ref{fig:collage}). Since the WISE and optical bands differ in both aperture size and pixel resolution, cross-wavelength color images are not feasible; instead, color combinations are restricted to within the optical or within the WISE bands (refer to Table \ref{Tab_2.2}).

To maintain FOV consistency, we integrate WISE data for only the two innermost apertures (see Figure \ref{fig:2.5}), preserving the 23x23 pixel data cube format. Although no additional spatial details are expected at this scale (see Sect. \ref{IP}), the WISE data still captures aperture flux in a format that enables direct cross-channel connections between optical and mid-infrared data at the image level.

Nevertheless, defected images can introduce challenges when generating color images (see bottom panel a) in Figure \ref{fig:collage}). Given that colors are derived from the ratio of images in different bands, the occurrence of unphysical negative or abnormally high/low values poses a significant concern. To address this issue, we examine whether the median value of neighbouring pixels is more than three times lower than the value of noisy pixels per band. If so, the central pixel's value is substituted with the median value of the surrounding pixels to smooth out fluctuations. In cases of non-detections or severely corrupted images where the largest value among all pixels in an image is < 0.001, the image is treated as a non-detection and all pixels are set to zero as default. After undergoing preprocessing, the images are utilized to create color images by exploring the six possible color combinations: $gr$, $gi$, $gz$, $ri$, $rz$ and $iz$. If either of the two flux images involved in creating a color image is identified as a non-detection, the resulting color image is set to a default value of -99.

At the catalog level, spatially invariant features, such as best-fit model classifications and signal-to-noise ratios (S/N), are processed separately in a dedicated channel, as they cannot be converted into image data. Notably, regarding normalization, we adopt a uniform min-max scaling approach to handle columns with cross-dependencies, such as flux bands. This strategy aims to preserve crucial information, such as the original shape of the SED.

Contrary to the conventional approach of stacking all available images into a single input \citep{hoyle2016, DIsanto2018, Pasquet2018, Dey2022, treyer2023cnn}, we find it advantageous to separate the color image data cube (23,23,24) from the flux band cutouts (23,23,32). This distinction is necessary because the color images often have differing pixel value scales, including negative values, which require specialized processing in our machine learning application, such as tailored loss functions in the CNN. Non-spatial attributes are then combined with image-based data at a subsequent stage of the model, allowing the model to capture both spatial and non-spatial aspects. By processing data types in parallel channels, we leverage their complementary information by merging the data at a later stage, enhancing the extraction of inter-band correlations and ultimately improving redshift precision \citep{ma2015multimodal,aitouahmed2023multimodality}. A detailed breakdown of the features integrated into each channel is provided in Table \ref{Tab_2.2}.


\section{Neural network} 
\label{sec5}

ML embodies an artificial intelligence paradigm where computers learn patterns and relationships from data, enabling them to make predictions or perform tasks without explicit programming. Multilayer perceptrons (MLPs), a feature-based feed-forward neural network, draw inspiration from their biological counterparts, namely excitable cells responsible for processing and transmitting information \citep{Rosenblatt, Goodfellow, MDeru}. Likewise, each assigned a distinct weight, computational input vectors can be organized into layers, relayed to one or more hidden layers, to compute a scalar output value \citep{Geron}. During training, these models learn data mappings by adjusting the weights and biases associated with their connections. The margin of change to the model after every training epoch is dictated by the choice of optimizer and loss function, which effectively calculates the Euclidean distance between the prediction and so-called ground truth in multi-dimensional feature space, thereby significantly impacting model convergence and performance.

The current state-of-the-art deep learning (DL) networks, characterized by their many hidden layers, have shown exceptional capabilities in handling complex non-linear tasks.

\subsection{Convolutional model layers}

Among such architectures, Convolutional Neural Networks \citep[CNNs;][]{fukushima:neocognitronbc,LeCun1989,lecuncnn1998} distinguish themselves by their remarkable effectiveness in handling grid-like data, a prevalent form of which is represented by images. CNNs leverage a model architecture that is particularly effective in tasks like image recognition, object detection, and image segmentation \citep{oshea2015introduction, Liu2022CVPR}. In convolutional layers, neurons establish connections exclusively with pixels within their receptive field, ensuring successive layers are linked only to specific regions of the previous layer. Subsequently, the model extracts low, image-level features in early layers and progressively complex, higher-level features in later layers. This allows the CNN to learn representations of the input data at multiple levels of abstraction. 

The convolutional operation involves sliding several kernels $K$, here of sizes 3$\times$3 or 5$\times$5 pixel, across the images with a fixed stride of size $s = 1$, compressing each mosaic element into a single scalar. This process entails element-wise multiplication, generating a set of $K$ feature maps. Filters, the learnable parameters of convolutional layers, enable the network to detect and highlight different aspects of the input data, such as edges or corners. Each convolution is followed by a pooling layer that reduces spatial dimensions and computational load. Pooling involves sliding a kernel, in our case of size 2 $\times$ 2 and $s = 1$, across the feature maps, selecting the maximum value (max pooling) within each kernel window. After flattening the data cube into a single array, it is passed through (several) fully connected (FC) layers. Each FC layer contains a layer-specific number of neurons $n$, tailored to the model's specific needs. The final FC layer's neuron count varies based on the task, with $n=1$ referring to regression tasks and $n\geq2$ to other endeavors such as multi-label classification. 

\subsection{Gaussian mixture models}
\label{sec_gmm}

By default, single output regression models \citep{Schuldt2021} provide point estimates without quantifying uncertainty. This severely limits their area of application, particularly in scenarios where quantifying the uncertainty is crucial, for example, when performing precision cosmology with Euclid \citep{Bordoloi2010, Scaramella2022,NewmanGruen2022}. By instead employing an architecture that provides PDFs, we can not only retrieve point estimates but also encapsulate the uncertainty associated with the predictions in a concise format \citep{DIsanto2018}. Given the inherent complexity in determining redshifts from only a few broadband cutouts, our results are expected to often exhibit degeneracy with multi-modal posteriors, making a single Gaussian insufficient for representing the photo-$z$ PDFs. Therefore, estimates are computed using Bayesian Gaussian Mixture Models \citep[GMMs, ][]{DudaHartStork01,viroli2017deep,Hatfield2020}. These networks provide a unique probabilistic modeling approach that differentiates them from traditional MLPs. One of the key distinctions lies in the output nature, where GMMs provide a set of variables to compute weighted multi-Gaussian distributions as opposed to single-point estimates. Subsequently, each component is characterized by its mean $\mu$, standard deviation $\sigma$, and weight $w$, allowing GMMs to produce a full PDF for a given set of inputs $x$. The PDF is expressed as
\begin{equation}
    P(x) = \sum_{k=1}^{K} w_{k} \cdot N(x | \mu_{k}, \sigma_{k}^{2}) \ ,
\end{equation}
with, $w_{k}$ representing the weight and $N(x| \mu_{k}, \sigma_{k}^{2})$ denoting the Gaussian distribution of the $k$-th component. As such, we extend our CNN approach by a GMM backend to output PDFs based on the information-rich feature maps produced during the front-end phase of the network. However, we encounter a limiting challenge with inputs of such small dimensions, i.e. 23 $\times$ 23, as they are not well-suited for established image-based Deep Learning architectures, such as "ResNet" \citep{he2015deep}, which typically require larger dimension scales, typically exceeding 200 $\times$ 200 pixels, to accommodate the large number of pooling layers they employ. Therefore, we developed a custom architecture to fit our data dimensionality. The resulting model architecture features roughly 490,000 trainable parameters, far fewer than found in comparable studies \citep[see][]{Pasquet2018, treyer2023cnn}, and is displayed in Figure \ref{fig:A.1} of Appendix \ref{A}.

\subsection{Model refinement}

Numerous hyper-parameters are crucial in shaping the network architecture while influencing training and convergence. Extensive optimization has been conducted across various parameters utilizing the Optuna framework \citep{akiba2019optuna}. The current configuration accounts for the vast array of potential combinations. The key parameters with the most significant impact are outlined below:

\begin{itemize}
    \item Batch size [8-2048]: The number of objects utilized in a single training iteration per epoch. Larger batch sizes offer computational efficiency, while smaller batches may help generalize better. 
    \item Learning rate [0.1-0.00001]: Controls the size of the model adjustment step during optimization. It influences the convergence speed and the risk of overshooting optimal settings. 
    \item Number of Gaussians [1-50]: Determines the complexity and flexibility of the GMM, influencing how well the model captures the underlying data distribution.
    \item Convolutional \& pooling layers [2-10]: the dimension of the kernel influences the size of the receptive field and the learnable features.
    \item Number of neurons and layers [10-300]: Determines the depth and complexity of the neural network, impacting its capacity to capture hierarchical features and relationships in the input data.
    \item Activation function: Introduces non-linearity to the model, enabling it to learn complex mappings between inputs and outputs, commonly a sigmoid, hyperbolic tangent (tanh), rectified linear unit (ReLu) or versions thereof.
    \item Dropout layer [0-0.9]: Refers to the fraction of randomly selected neurons that are temporarily dropped or ignored during training, helping to prevent overfitting by enhancing network robustness and generalization.
    \item  Max and average pooling: Pooling layers, such as Max Pooling and Average Pooling, are used to downsample the spatial dimensions of the input feature maps, reducing the computational load. 
    \item Batch normalization: Batch Normalization can improve the training stability and speed of convergence in neural networks by subtracting the batch mean and dividing by the batch standard deviation.
\end{itemize}

\subsection{Loss functions}

When utilizing PDFs instead of point estimates, it is crucial to quantify whether a predicted PDF effectively reflects the ground truth, in our supervised ML case, spec-$z$ \citep{Sanchez2014,Malz2018, Schmidt2020,treyer2023cnn}. The PDF predicted by a model should be concentrated near the true value. A straight-forward and meaningful proper scoring rule (i.e. one which is lowest when the prediction is at the truth), is the product of probabilities at the true redshifts. The negative log-likelihood (NLL)\footnote{Transforming probabilities into their logarithms simplifies the mathematical operations involved (multiplication into addition), rendering the function more amenable to optimization techniques such as gradient descent, facilitating the search for optimal model parameters.} of which passed to a minimizer

\begin{equation}
    E(w) = - \sum_{n=1}^{N}\textrm{ln}\bigg(\sum_{k=1}^{k}w_{k}(x_{n})N(y|\mu_{k}(x_{n}),\sigma_{k}(x_{n})^{2})\bigg) \ ,
\end{equation}
yields the likelihood of observing a data distribution given a specific set of model parameters. This expression coincides with the average Kullback-Leibler divergence \citep[KL, ][]{KL} when going from a delta PDF at spec-$z$ to the photo-z PDF. 

An alternative growing in popularity is given by the Continuous Ranked Probability Score (CRPS), initially applied in weather forecasting \citep{Grimit2006}, which serves as a valuable metric for photo-z estimation via PDF quantification \citep{DIsanto2018}. Computationally, CRPS calculates the integral of the squared difference between the predicted cumulative probability distribution function (CDF) and a Heaviside step-function $H$($x$) at the value of the spec-$z$ ($x_{z}$) as

\begin{equation}
    \textrm{CRPS}(F, x_{z}) = \int_{-\infty}^{\infty} \bigg[\int _{-\infty}^{x} f(t)dt - H(x-x_{z})\bigg]^{2}\textrm{d}x \ .
\end{equation}
For a finite mixture of normal distributions $M$, the CRPS can be expressed in closed form:

\begin{equation}
    \textrm{CRPS} = \sum_{i=1}^{M}w_{i}A(\delta_{i},\sigma_{i}^{2}) - \frac{1}{2} \sum_{i=1}^{M}\sum_{j=1}^{M}w_{i}w_{j}A(\mu_{i}-\mu_{j}, \sigma_{i}^{2}+\sigma_{j}^{2})\ ,
\end{equation}

\noindent with $\delta_{i}=y-\mu_{i}$, the uncertainty $\sigma_{i}$ and weight $w_{i}$ of the $i$-th component respectively, as well as

\begin{equation}
\begin{split}
    A(\mu,\sigma^{2}) =  \mu \bigg[2\Phi \bigg(\frac{\mu}{\sigma}\bigg)-1\bigg] + 2\sigma\phi\bigg(\frac{\mu}{\sigma}\bigg)\ , \\
    \phi (x) = \frac{1}{\sqrt{2\pi}} \textrm{exp}\bigg(-\frac{x^{2}}{2}\bigg)\ , \\
    \Phi(x) = \int_{-\infty}^{x} \phi(t) \textrm{d}t \ .
\end{split}
\end{equation}

By extending the CRPS loss via normalizing it by (1+z), we adjust the penalty for prediction errors based on redshift. In doing so, we prioritize accuracy for low and intermediate redshift sources by imposing higher constraints while allowing for more leniency in less critical high-redshift areas. Integrating over the entire range of possible redshift values, the CRPS takes into account both location and spread of the predicted PDF. It, therefore, provides a comprehensive evaluation metric, generating globally well-calibrated PDFs \citep{dey2022recalibrating}.

\subsection{Training \& data augmentation}

\begin{figure}[b!]
\centering
\includegraphics[width=9cm]{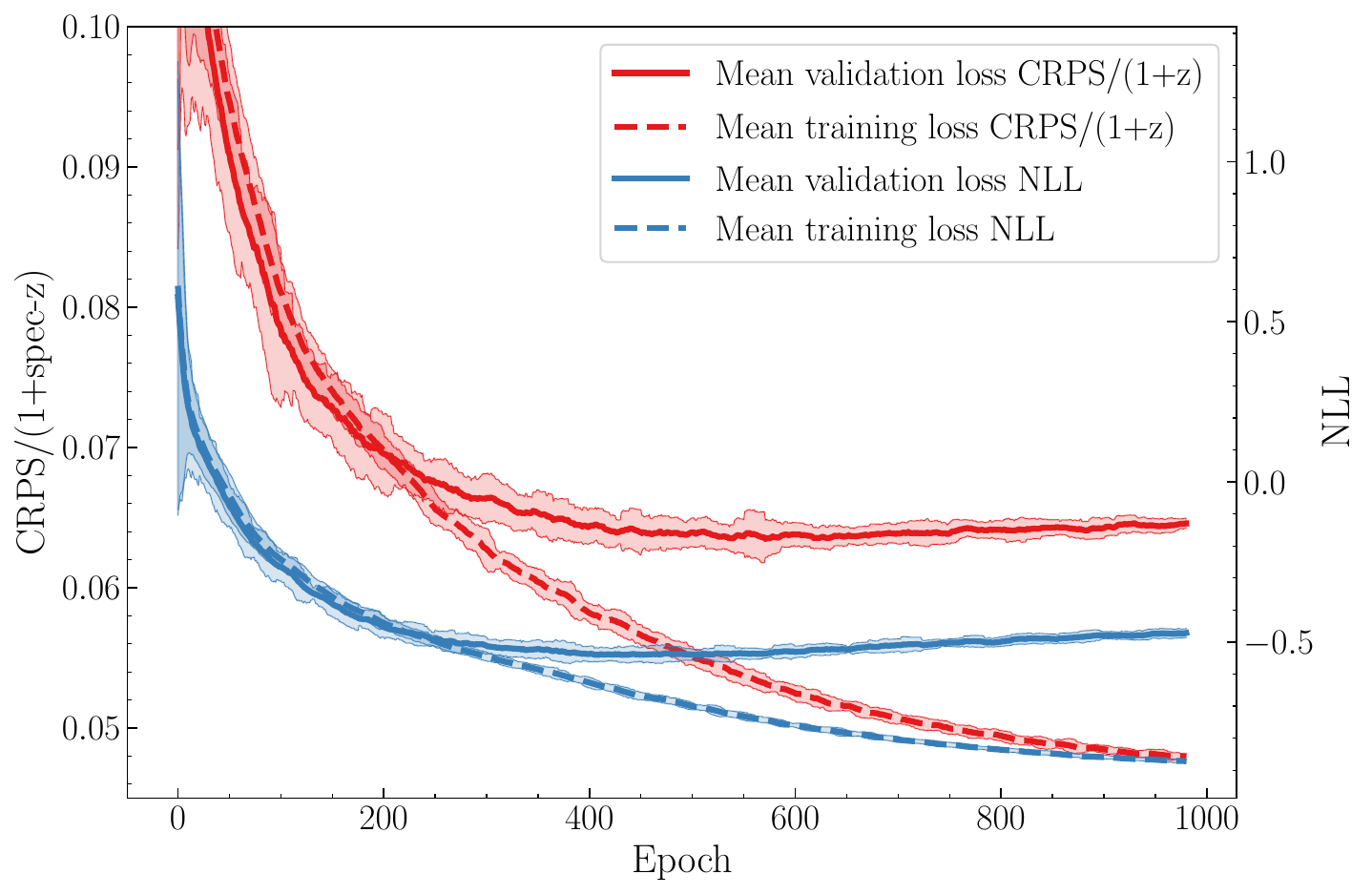}
\caption{Training (dashed) and Validation (solid) loss curves for a single {\sc{PICZL}} model. The solid or dashed lines represent the mean loss values computed over a window size of 20, while the shaded areas denote the 1$\sigma$ error range around the mean. The NLL loss is depicted in blue, while the normalized CRPS loss is shown in red.}

\label{fig:3.2}
\end{figure}

Our dataset is divided into training and validation sets in an 80:20 ratio. The model is trained over 1000 training epochs utilizing the Adam algorithm \citep{kingma2017adam} along with a learning rate scheduler to adjust the learning rates during training dynamically. The inclusion of both is a consequence of the Optuna hyperparameter optimization. Typically, the model achieves its lowest validation loss around the 600th epoch when considering both loss functions (see Figure \ref{fig:3.2}). After this point, although the training loss continues to decrease, the validation loss begins to increase, indicating overfitting. This overfitting likely arises due to the limited size of our training sample, allowing the model to memorize specific patterns rather than learning generalizable features. We implement a checkpoint system to mitigate this, saving the model's architecture and parameters when it reaches its lowest validation loss. This approach ensures that we capture the best-performing model configuration while preventing overfitting.

To further exploit the advantages of deep learning, particularly its capacity to perform well with extensive datasets, we have tested the incorporation of data augmentation techniques into our methodology. Recognizing the substantial impact of training data amount on model performance, we employ common augmentation data products, such as rotated and mirrored images. Although these transformations significantly increase the apparent size of the dataset, we do not observe an increase in performance when including augmentation techniques \citep{Dey2022,treyer2023cnn}. This is potentially not surprising as the influence of augmentation techniques depends on several factors. While CNNs are inherently equivariant to translations within images, enabling them to recognize objects even if they are shifted or off-centered, they and most other machine-learning approaches are not naturally equivariant to rotations and reflections. This lack of rotational equivariance means that the model output effectively depends on the rotation of the input images. However, as most of our inputs are approximately point sources with radial symmetry, we do not explore this aspect further in this study. More recent publications have focused on developing rotationally equivariant CNN architectures, which have shown improved performance despite increased computational costs \citep{cohen2016,weiler2021,Natalie2024}. Incorporating such architectures could be particularly beneficial for future works dealing with large amounts of imaging data.

\subsection{Model ensemble}
\label{Model-Ensemble}

Ensemble models capitalize on the diversity of multiple models to improve prediction accuracy. Our approach involves diversifying models by adjusting parameters such as the number of Gaussians per GMM, learning rates, batch sizes, and the choice of loss function. We employ both NLL and CRPS as complementary loss functions to achieve this. While CRPS excels in achieving lower outlier fractions, NLL enhances overall variance. Each loss function trains a set of 144 models, resulting in a diverse pool of 288. In line with \cite{treyer2023cnn}, we achieve superior performance by randomly combining equally weighted CRPS and NLL models from the pool, compared to the best-performing individual model \citep{Dahlen2013,NewmanGruen2022}. This finding is also aided by the result that roughly 30\% of outliers are model-specific as opposed to 20\% of outliers appearing in all models, therefore considered to be genuine. Put differently, 80\% of outliers were found to not be an outlier in at least one other model. 

\begin{figure*}[t!]
\centering
\includegraphics[width=19cm]{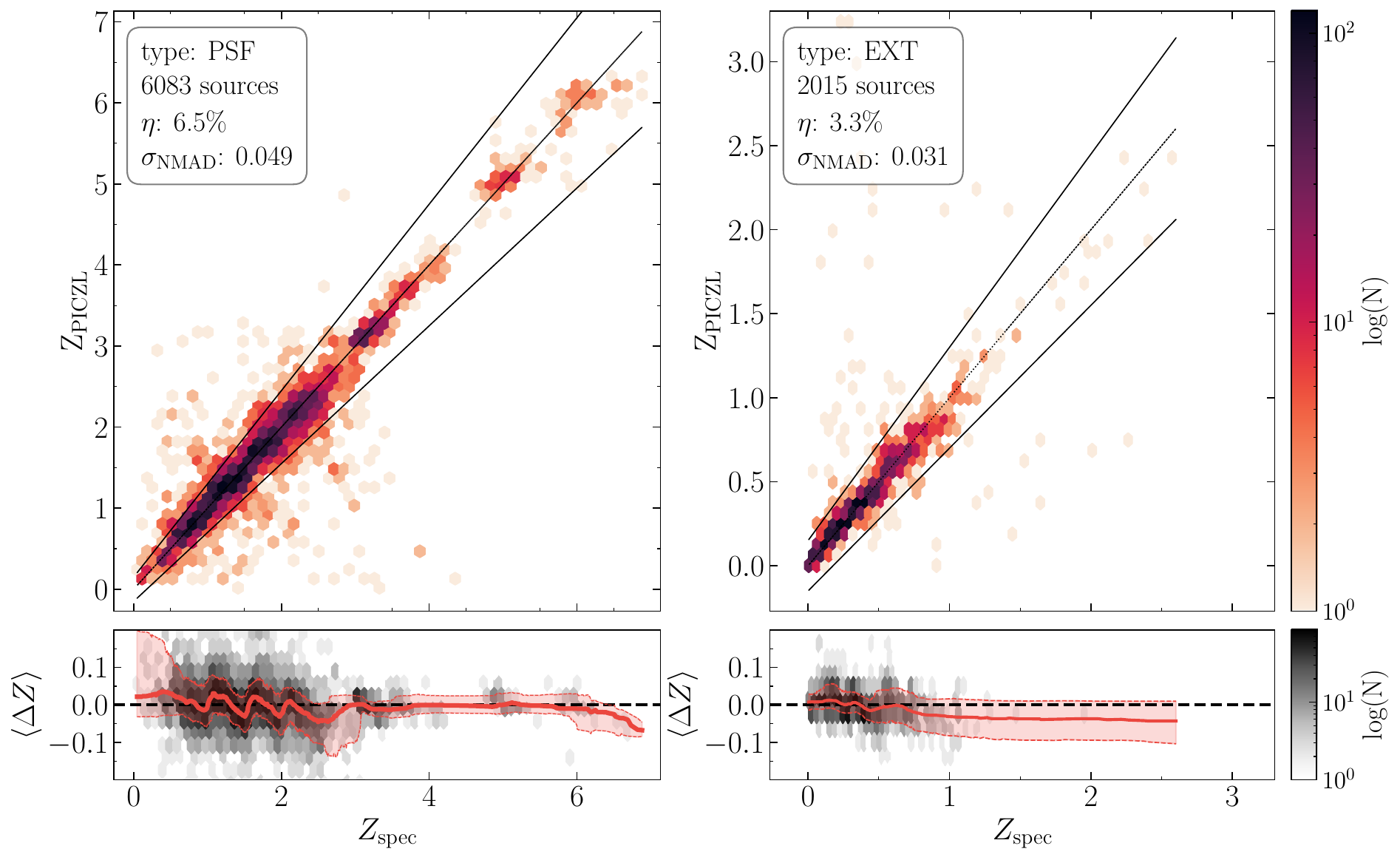}
\caption{Binned Scatter plots of photo-z obtained from our model Z$_{\mathrm{PICZL}}$ versus spectroscopic redshift Z$_{\mathrm{spec}}$, for sources classified as type point-like (PSF, left) and extended (EXT, right). Each plot includes identity and two lines denoting the outlier boundary of $ \frac{|z_{\mathrm{phot}} - z_{\mathrm{spec}}|}{(1+z_{\mathrm{spec}})} > 0.15$. Normalized residuals and trend lines and errors are shown in the bottom panels, aiding in the visualization of systematic deviations across the redshift range.}

\label{fig:4.4}
\end{figure*}

We additionally recognize that the initial configuration of each model has a minor influence on the performance. Although training additional models could further enhance our results, the computational resources required to generate three to five sets of models would make this approach impractical relative to the potential performance gains. To create an effective ensemble, we evaluate the outlier fraction of all individual models on an unseen test sample. From this evaluation, we select models based on their relative influence on the model performance. We then observe how the ensemble's performance evolves as more models are continuously incorporated. Our ensemble ultimately comprises 10 models or 84 Gaussians, evenly incorporating models optimized using both NLL and CRPS approaches. To refine the model weights within the ensemble, we employ Optuna for fine-tuning. Subsequently, the ensemble posterior likelihood distribution $P_{\mathrm{ensemble}}$(x) is given by:

\begin{equation}
 P_{\mathrm{ensemble}}(x) = \sum_{i=1}^{N} w_{i}P(i)(x) \ ,
\end{equation}

\noindent where the weights $w_{i}$ are normalized such that:

\begin{equation}
    \sum_{i=1}^{N} w_{i} = 1 \ ,
\end{equation}

\noindent to assure that the integral of the ensemble PDF is equal to 1. We obtain a point estimate for our photo-$z$ prediction by identifying the dominant mode of the resulting PDF, recognizing that the mean or median could be misleading for highly non-Gaussian and bi- or multi-modal distributions. Additionally, we provide asymmetric 1 and 3 sigma upper and lower errors by evaluating the PDF within the 16th/84th and 0.15/99.85th percentiles, respectively, along with the entire PDF.


\section{Photo-z results} 
\label{sec6}

We evaluate the performance of {\sc PICLZ} on both point estimates and PDFs. This comprehensive assessment includes testing the statistical reliability and accuracy of our predictions. We adopt several commonly employed statistical metrics, providing insights into the accuracy, precision, and reliability of the photo-z estimates. In line with the definitions from the literature, we use the following metrics:

\begin{itemize}
    \item Prediction bias: the mean of the normalised residuals, $\langle \Delta z \rangle  = \frac{(z_{\mathrm{spec}} - z_{\mathrm{phot}})}{(1+z_{\mathrm{spec}})}$.
    \item Variance: following \cite{Ilbert2006}, we adopt the standard deviation from the normalised median absolute deviation $\sigma_{\mathrm{NMAD}} = 1.4826 \times \mathrm{Median}  \frac{|z_{\mathrm{spec}} - z_{\mathrm{phot}}|}{(1+z_{\mathrm{spec}})}$.
    \item Fraction of outliers $\eta$ is determined by the proportion of photo-z estimates with absolute normalised residuals exceeding $\eta$ = $ \frac{|z_{\mathrm{spec}} - z_{\mathrm{phot}}|}{(1+z_{\mathrm{spec}})} > 0.15$. 
    \item PIT score: a diagnostic tool used to evaluate the calibration of probabilistic forecasts by evaluating the cumulative distribution function (CDF) given the PDF at the (true) spec-$z$ $z_{\mathrm{corr}}$ as 
    $\mathrm{CDF}(z_{\mathrm{corr}}) = \int_{0}^{z_{\mathrm{corr}}} \mathrm{PDF}(z) \, dz$.

\end{itemize}

\noindent We quantify {\sc{PICZL}}'s performances on a validation sample of 8098 sources and find an outlier fraction $\eta$, of 5.6\% with a variance $\sigma_{\mathrm{NMAD}}$, of 0.045. Since the validation sample is used for feedback in training, it should be mentioned that the results mentioned above are likely overestimating the performance and therefore may not representative of what a future user could achieve with say, a test set previously unknown to the model, leave-one-out, or k-fold cross-validation. To this end, we have estimated our performance on independent, blind, X-ray-selected samples (see Sect. \ref{blind} and \ref{sec9}).

\begin{figure}[h!]
\centering
\includegraphics[width=8.5cm]{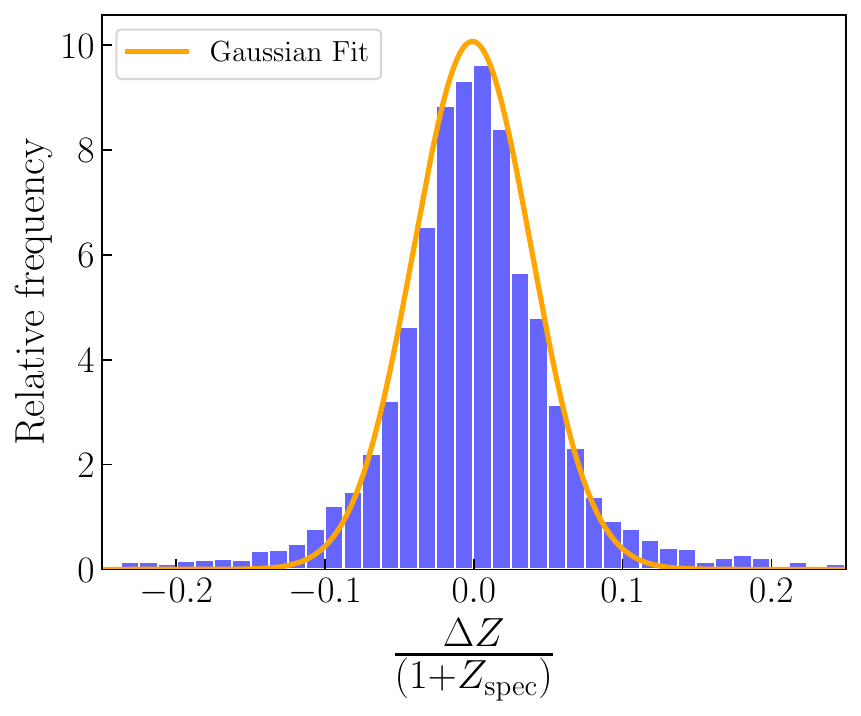}
\caption{Histogram distribution of the normalised residuals, overlaid with a Gaussian fit (orange). The parameters of the Gaussian distribution are determined by the prediction bias and $\sigma_{\mathrm{NMAD}}$ values obtained from Table \ref{tab:accuracyBreackdown} for the validation sample.}

\label{fig:4.2}
\end{figure}

Figure \ref{fig:4.4} compares photometric versus spectroscopic redshifts, split by point-like (type=PSF) and extended (type=EXT) morphology. With respect to comparable work \citep[e.g., Figure 13 from][]{Salvato22}, we observe enhanced performance with a much-reduced fraction of outliers, especially for PSF sources. These objects lack morphological information at redshifts $z \gtrsim 1$, hence we attribute this improvement to the model's ability to recognise how the radial extension of the azimuth profile of sources changes with redshift (refer to Figure \ref{fig:2.4}). The scatter, for both PSF and EXT distributions, is tight and symmetrically distributed around the $z_{\mathrm{{\sc{PICZL}}}} = z_{\mathrm{spec}}$ identity line, with outliers appearing randomly scattered, suggesting stable performance across the redshift range and minimal systematic errors (see Figure~\ref{fig:4.2} and \citet{Dey2019} for reference).

Given that our point estimates are derived from PDFs, we must ensure their global calibration and accuracy. To achieve this, we employ the probability integral transform (PIT) statistic \citep{Dawid1984,Gneiting2005}, a widely accepted method in the field for assessing the quality of redshift PDFs \citep{Pasquet2018,Schuldt2021,NewmanGruen2022}. An ideal scenario is represented by a uniformly distributed histogram of PIT values. Any deviation from uniformity can indicate issues in PDF calibration. Under-dispersion may suggest overly narrow PDFs, while over-dispersion often results from excessively wide PDFs \citep{DIsanto2018}. Peaks close to zero or one can be explained by catastrophic outliers, where the true redshift lies so far in the wing of the PDF that it essentially falls outside of it.

\begin{figure}[t!]
\centering
\includegraphics[width=9cm]{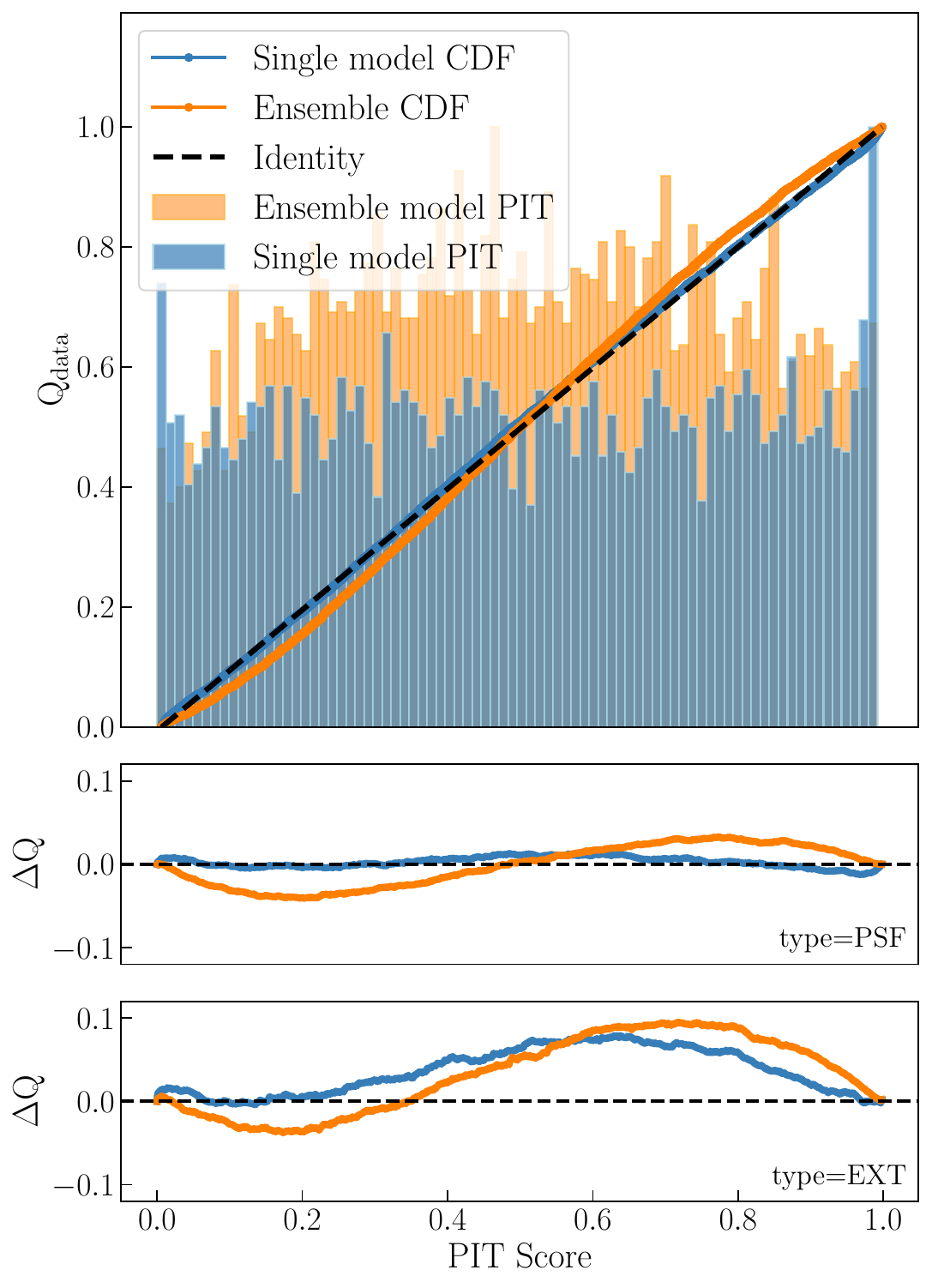}
\caption{The top panel displays the QQ plot, comparing identity (flat histogram) against the PIT values derived from our redshift PDFs of the validation sample. The bottom panel shows the differences between the QQ plot and identity, highlighting systematic biases or trends for the two morphological classes (PSF and EXT). For all panels, blue refers to results achieved with a single model (see Figure \ref{fig:A.1}) while orange reflects the results obtained when utilizing ensemble results (refer to Sect. \ref{Model-Ensemble})}.

\label{fig:4.3}
\end{figure}

We employ the PIT histogram (top panel Figure \ref{fig:4.3}) alongside quantile-quantile (QQ) plots (bottom panel Figure \ref{fig:4.3}), to visually assess the statistical properties of our model predictions. For reference, these can be compared to Figure 2 of in \citet{Schmidt2020}; however, those code performances are completely dominated by inactive galaxies, making a direct comparison impractical. The QQ plot compares the CDFs of observed PIT values and identity $U(0,1)$, to visualize QQ differences split by morphological type.

\begin{table*}
\centering
\caption{Overview of wall-clock time $\tau$, $\Delta z$, $\sigma_{\textrm{NMAD}}$ and $\eta$ for sources classified as being extended (EXT) or point-like (PSF) from a single model (as opposed to an ensemble utilized for the results depicted in Figure \ref{fig:4.4}) trained with four input configurations.}

\begin{tabular}{l|c|c|c|c|c|c|c}

Features & $\tau$ & $\langle\Delta z\rangle_{\mathrm{EXT}}$  &  $\langle\Delta z\rangle_{\mathrm{PSF}}$ &  $\sigma_{\textrm{EXT}}$   & $\sigma_{\textrm{PSF}}$  & $\eta_{\mathrm{EXT}}$ &  $\eta_{\mathrm{PSF}}$\\
\hline
\hline
&\\ 
model flux\textsuperscript{{\textrm{a}}}  &11:14,41&-0.045&-0.043&0.062&0.096&  11.1\% &  29.1\% \\
model flux + ap. flux\textsuperscript{{\textrm{b}}}  &16:03,82&-0.020&-0.012&0.044&0.058& 5.4\%& 10.2\%\\
model flux + ap. flux + images\textsuperscript{{\textrm{c}}} & 20:39,91 & -0.005 & -0.002& 0.045& 0.052& 4.3\%& 8.4\%\\
images ($griz$ + $W1-W4$ ap. flux)\textsuperscript{{\textrm{d}}} &17:17,07&-0.029&-0.003&0.043&0.059& 5.2\% & 9.6\%\\
&\\ 
\hline

\end{tabular}

\tablefoot{The last two setups are an improvement over the previous performances, with the last setup providing comparatively good results, despite ignoring the features provided by catalogs, particularly relevant for potential implementation in e.g., LSST.\\
\tablefoottext{a}{Catalog-based model fluxes only, as traditionally done;}
\tablefoottext{b}{Catalog-based aperture fluxes and respective colors as done in \citet{saxena2024};}
\tablefoottext{c}{Current {\sc PICZL} setup as per Table \ref{Tab_2.2};}
\tablefoottext{d}{{$griz$ + $W1-W4$ aperture flux images.}
}}


\label{tab:loss}
\end{table*}

A well-calibrated model will exhibit a QQ plot closely following identity with a flat PIT histogram, indicating accurate and reliable redshift predictions. Our analysis shows that, while asymmetry in the PIT distribution could suggest systematic bias, the QQ plot reveals small residuals overall. Notably, for both the single- (refer to Sect. \ref{sec_gmm} and Figure \ref{fig:A.1}) and ensemble model (refer to Sect. \ref{Model-Ensemble}), the curves do not deviate significantly from zero, lesser so for PSF-type than EXT-type objects, indicating well-calibrated models. A distinct observation is the shift towards central PIT values for the ensemble model. This shift is not unexpected, given that the ensemble approach integrates multiple redshift solutions, each potentially contributing different peaks to the resulting ensemble PDF. Consequently, the ensemble PDF exhibits a broader bulk of probability across redshift, with PIT scores accumulating more area under the curve when integrated to the main mode. As a result, rather than yielding a flat PIT distribution, the histogram shifts towards having a concentration of values around 0.5. While this phenomenon aligns with our expectations, further enhancements may be realized by incorporating metrics tailored instead to local calibration accounting for population-specific subgroups \citep[see e.g.][]{zhao2021diagnostics,dey2022recalibrating}.

\subsection{Model performance for different inputs}

Images provide a more comprehensive view of astronomical sources by capturing their full spatial structure and light distribution, offering finer details on morphology, apparent size and extended features that are often lost in the averaging process of aperture photometry. This is pivotal, as SED features crucial for determining redshift solutions and resolving degeneracies mostly reside within the host galaxy \citep{Soo2017, Wilson2020}. In particular for AGN, images help to spatially separate the pixels corresponding to AGN-dominated emission and those corresponding mostly to the host galaxies. This approach enhances our ability to isolate and analyze individual pixel strings across multiple filters, to effectively construct SEDs for the host galaxies independently. We thereby capture subtle features, neighbors, and patterns that may not be discernible from total flux measures. Critically, the independent photo-z from the AGN and host must agree, thus narrowing the overall source PDF.

Table \ref{tab:loss} presents the fraction of outliers for both EXT and PSF sources using different configurations of the {\sc{PICZL}} algorithm. The first setup replicates the feature-only method, relying solely on total fluxes from the catalog, which, as seen in other studies, results in the fraction of outliers for PSF sources being nearly three times higher than for EXT sources \citep{Salvato22}. In the second configuration, we adopt the approach of \citet{saxena2024}, incorporating the 2D light distribution and color gradients within annuli. The third configuration, i.e., {\sc{PICZL}} as outlined above, enhances this by incorporating images, supplemented with catalog values transformed into image format, which leads to an additional improvement over the method in \citet{saxena2024}.

Since images inherently capture all features typically extracted numerically and presented in catalogs, we show in the fourth case that {\sc{PICZL}} achieves excellent results without feature/catalog information, using solely optical images supplemented by WISE aperture flux maps, even without explicit hyperparameter tuning. This approach is particularly promising for future surveys such as LSST and Euclid, where relying solely on multi-band imaging, including NIR/IR, will eliminate the need for complex source modeling or aperture photometry, regardless of the sources being galaxies or AGN.

\begin{figure}[b!]
\centering
\includegraphics[width=9.1cm]{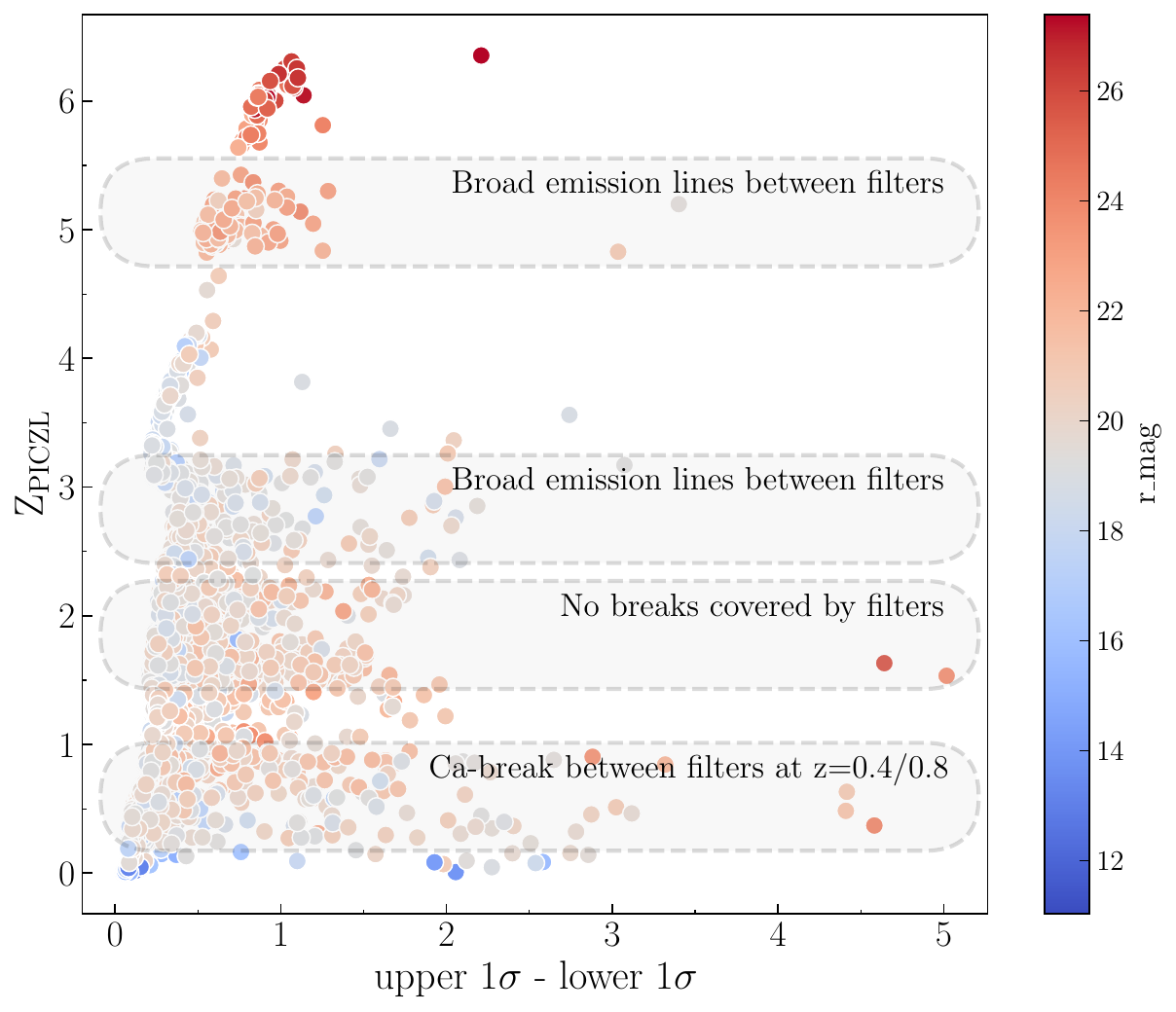}
\caption{{\sc{PICZL}} point estimates as a function of PDF width, represented by the difference between upper and lower 1$\sigma$ values. Each data point is color-coded according to its $r$-band magnitude from the LS10. The tilt of the distribution traces the increased uncertainties for PDFs of fainter sources. Horizontal distributions represent areas of large uncertainty in the photo-z, indicating that {\sc PICZL} errors are realistic.}

\label{fig:4.7}
\end{figure}

\begin{figure*}[t]
\centering
\includegraphics[width=19cm]{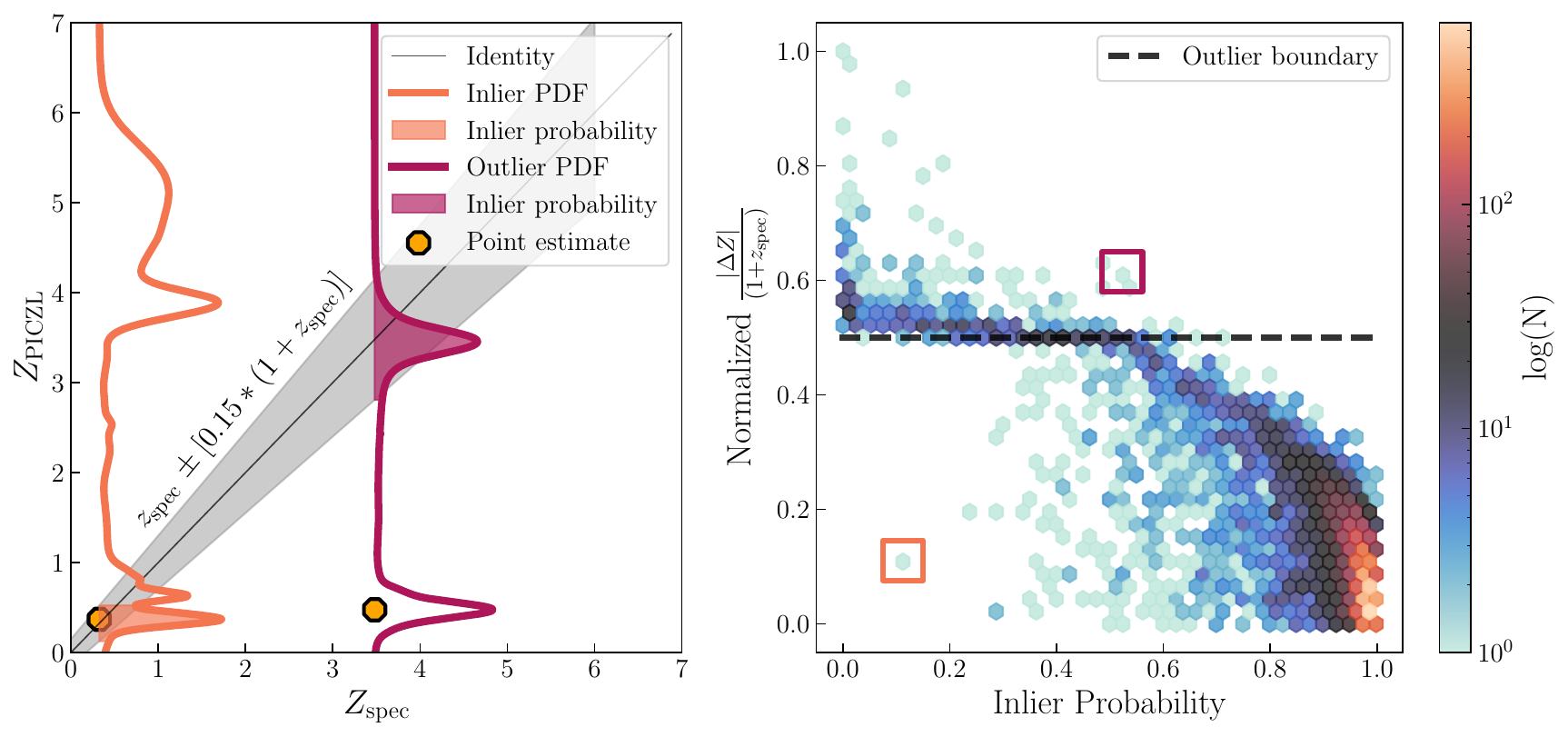}
\caption{Left: Examples of an inlier (orange) and an outlier (purple) PDF. By definition, the point estimate derived for the two sources from the main mode falls within or outside the grey shaded region. Both sources show a secondary peak in the PDF. We can define the probability of being an inlier (colored area under the PDF which coincides with the grey shaded area).
Right: Normalized $\frac{|\Delta Z|}{(1+z_{\mathrm{spec}})}$ as a function of inlier probability, color-coded by number density. 
The majority of the inliers have a high inlier probability, indicating the stability of our results (see main text for details).}

\label{fig:5.2}
\end{figure*}

\subsection{Blind sample comparison}
\label{blind}

To evaluate the robustness of our approach, we tested {\sc {PICZL}} on an independent blind sample from the Chandra Source Catalog 2 (CSC2\footnote{\url{https://cxc.cfa.harvard.edu/csc/}}). This sample, selected to ensure no overlap with the training data, but yield a comparable X-ray to MIR distribution, is the same one used in the \textsc{CircleZ} analysis \citep{saxena2024}, allowing for a direct comparison with their results. After filtering for sources with spectroscopic redshift and removing duplicates with the training dataset, the sample comprises 416 sources within the LS10-South area. For comparison, \cite{saxena2024} achieve an outlier fraction, $\eta$, of 12.3\% and a normalized median absolute deviation, $\sigma_{\mathrm{NMAD}}$, of 0.055, while our algorithm yields a superior $\sigma_{\mathrm{NMAD}}$ of 0.046 and $\eta$ of 8.3\% (see Table \ref{tab:accuracyBreackdown} and Figure \ref{fig:F}). This comparison, therefore, demonstrates how utilizing image cubes further improves the already good results achieved using 2D-information from a catalog. For more details on the sample, see \citet{saxena2024}.

\subsection{Prediction uncertainty quantification}

Accurate error estimation is crucial for assessing the reliability of any prediction, particularly in those astrophysical contexts where uncertainties can significantly impact the interpretation of data. To this end, we provide asymmetrical 1$\sigma$ and 3$\sigma$ errors for every source, together with the photo-$z$ PDF, as detailed in Sect. \ref{Model-Ensemble}. These error estimates offer a comprehensive understanding of the potential variance in our predictions, accurately reflecting the inherent uncertainties in our data and model. We include asymmetrical errors, as the true distribution of uncertainties is often not symmetrical. This asymmetry arises from factors such as photometric noise and degeneracies in the highly non-linear color-redshift space, where certain higher or lower redshift values may be more probable than their counterparts, leading to skewed PDFs. Although symmetrical redshift errors are observed for the majority of sources, approximately one-third of the sources exhibit asymmetrical 1$\sigma$ errors with deviations in redshift up to $|\Delta z_{\mathrm{asym}}| \simeq 0.25$. 

Figure \ref{fig:4.7} displays {\sc{PICZL}} point estimates as a function of PDF width, represented by the upper 1$\sigma$ minus lower 1$\sigma$, color-coded by the Legacy Survey $r$-band magnitude, as it offers the most extensive coverage compared to the $g$, $i$, or $z$ bands, containing the largest number of sources. We observe wider PDFs correlating with higher redshifts and fainter sources characterized by higher photometric errors or lower S/N, as discussed in Sect. \ref{Ph-errors}. Additionally, the widths of the PDFs exhibit noticeable horizontal structures, largely influenced by the nature of the LS10 filters. We find that PDFs with wider modes, indicating less certain redshift estimates, often correspond to redshift ranges where key spectral features, such as the Ca break, fall outside the filter coverage (e.g. $g \& r$ at $z\simeq0.4$, $r \& i$ at $z\simeq0.8$ and $z \simeq$ 1.5 or $z \simeq$ 2.2). At redshifts approaching $z \approx 3$ and extending below $z \approx 5$, the Lyman-$\alpha$ break starts to fall within the filter ranges, which contributes to narrower PDFs with increased PDF width only observed for extremely faint sources at even higher redshifts. Lastly, we find that the number of sources with wider PDFs increases as a function of $\frac{|\Delta Z|}{(1+Z_{\mathrm{spec}})}$, indicating that narrower PDFs correspond to more accurate point predictions.

\subsection{Insights from multimodal PDFs}

To investigate whether predicted outliers are genuine anomalies or instead stem from machine learning processes, we inquire whether secondary peaks in PDFs are physically meaningful or training artifacts. For each source, we compute the fraction of its PDF, that satisfies the condition $\frac{|\Delta Z|}{(1+z_{\mathrm{spec}})} \geq 0.15$, to provide the probability of being considered an inlier. The left panel of Figure \ref{fig:5.2} illustrates this with examples of two PDFs (one for an inlier and one for an outlier), where we highlight the corresponding inlier probabilities. In the right panel of Figure \ref{fig:5.2}, we plot $\frac{|\Delta Z|}{(1+z_{\mathrm{spec}})}$ re-normalised to scale [0,1], such that

\begin{equation*}
\frac{|\Delta z|}{(1+z_{\textrm{spec}})}_{\textrm{norm}} = \begin{cases}
\frac{|\Delta z|}{(1+z_{\textrm{spec}})} \cdot \frac{0.5}{0.15} &\text{if} \quad \frac{|\Delta z|}{(1+z_{\textrm{spec}})} < 0.15\\
\frac{|\Delta z|}{(1+z_{\textrm{spec}})} \cdot \frac{0.5}{0.15} + 0.5 &\text{if} \quad \frac{|\Delta z|}{(1+z_{\textrm{spec}})} \geq 0.15
\end{cases}
\end{equation*}

\noindent as a function of inlier probability for all sources. The horizontal dashed black line separates the inliers (< 0.5) from the outliers ($\geq$ 0.5). The majority of inliers are concentrated at high inlier probabilities. This suggests that most inliers provide confident, unimodal PDFs with low errors. Likewise, most outliers have low inlier probability, with only a few outliers having more than 50\% inlier probability. While this quantity cannot be computed for sources lacking spectroscopic redshift, it can be used to evaluate the reliability of our point estimates considering their associated PDFs.

\begin{figure}[b!]
\centering
\includegraphics[width=9cm]{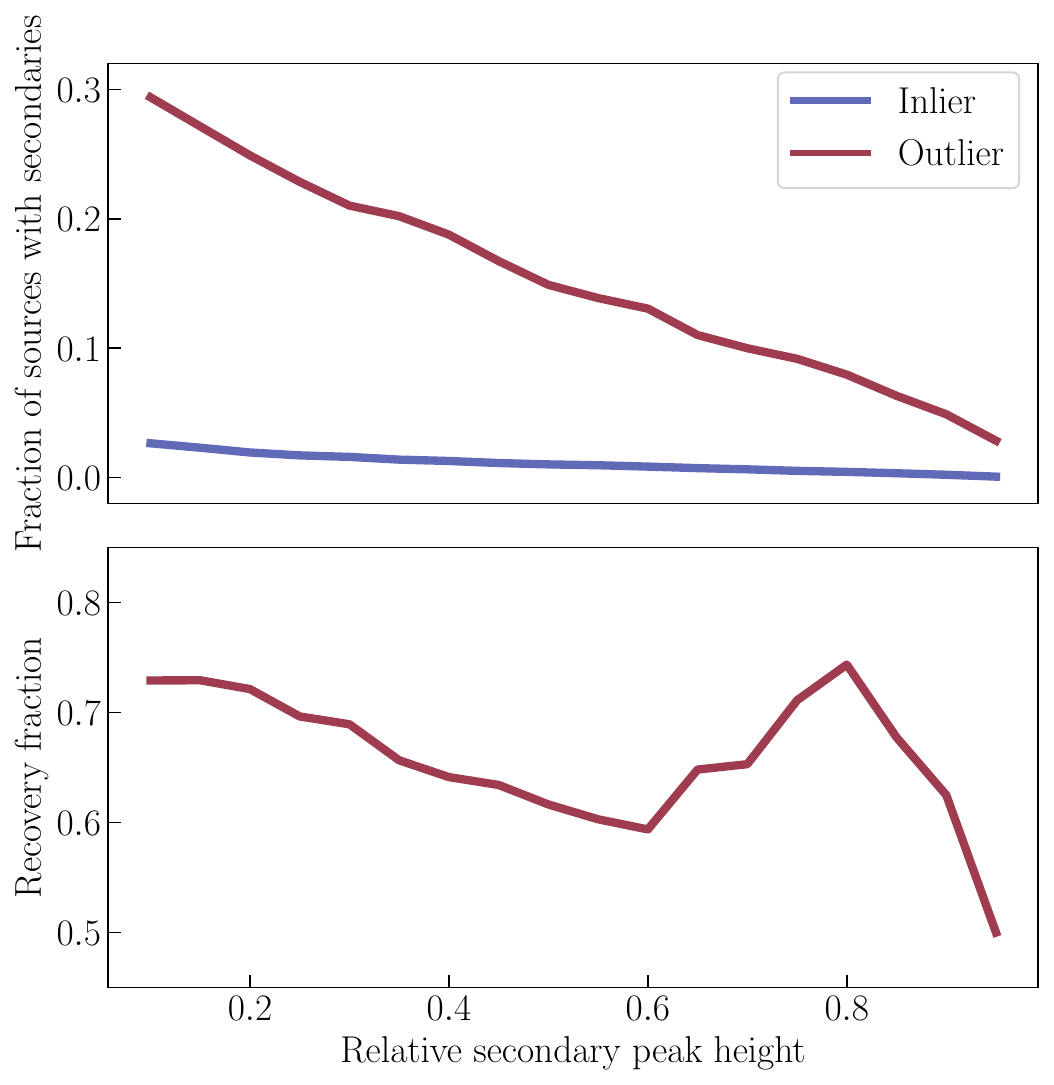}
\caption{Top: Fraction of sources having a PDFs presenting secondary peaks for outliers (burgundy) and inliers (blue). Bottom: Corresponding recovery fraction for outliers. Both metrics are plotted as a function of the relative prominence threshold of the secondary peak.}
\label{fig:5.1}
\end{figure}

To evaluate whether ensemble solutions with broad or complex distributions effectively capture the true redshift, we identify sources with multiple peaks and measure the proportion of PDFs exhibiting more than one mode. For these sources, we define a secondary peak as significant if it accounts for at least 10\% of the height of the primary peak, which is the minimum prominence threshold used in our analysis. The top panel of Figure \ref{fig:5.1} shows that only 3\% of inliers exhibit secondary peaks as opposed to the roughly 30\% of outliers, where occurrences of strong secondary modes decrease with increasing prominence for both cases. While a single mode typically indicates a secure estimate for inliers, the presence of a unique mode alone can subsequently not be used to determine whether a source is an outlier.

\noindent Conversely, in the bottom panel of Figure \ref{fig:5.1}, we present the recovery fraction, which denotes how often a secondary peak in PDFs exhibiting multi-modal distributions corresponds to the true redshift. Among the outliers with 30\% multi-modal PDF, more than 70\% have one of their secondary peaks corresponding to $\frac{|\Delta Z|}{(1+Z_{\mathrm{spec}})} \leq 0.15$, indicating that there is significant probability that the redshift could consequently be considered as an inlier if the peak heights were reversed. Given the PIT histograms shown in Figure \ref{fig:4.3}, it appears likely that the the PDFs are accurately capturing the relative frequency of multi-\& bimodal PDFs, as if they were not, there would likely be bias evident in the overall PIT distribution. In other words, using this particular dataset, despite incorporating full multiwavelength image cutouts and catalog-level information, there persist areas of parameter space that are legitimately degenerate in redshift, and the PDF parameterization looks to be capturing that degeneracy accurately. Consequently, we recommend using the entire PDF rather than point estimates, when possible.


\section{{\sc PICZL} applied to other surveys}
\label{sec9}

We want to investigate how {\sc PICZL}, utilizing LS10 photometry and imaging, performs in determining photo-z for AGN in a more generic setting. Our focus is set on the LSST deep drilling fields (DDFs), selected to study SMBH growth across the full range of cosmic environments. These fields offer deeper, more comprehensive spectroscopic redshift coverage, along with a broader range of high-sensitivity bands, providing an enhanced dataset for photo-z estimation.

\subsection{XMM-SERVS}

\begin{table*}
\centering
\caption{Summary of the XMM-SERVS samples and their respective photo-$z$ metrics.}  
\begin{tabular}{lccccc||cccccc}

Sample & size & w/ LS10  & w/ spec-z & blind & w/ photo-z$_{\textrm{S}}$=-99 & $\langle \Delta z \rangle$$_{\textrm{S}}$ & $\langle \Delta z \rangle$$_{\textrm{P}}$ & $\sigma$$_{\textrm{S}}$ & $\sigma$$_{\textrm{P}}$ & $\eta$$_{\textrm{S}}$  & $\eta$$_{\textrm{P}}$ \\
     &        & optical CTPs &  & &  & & & & & & \\ 
\hline
\hline
& & &  & &  & & & && &\\ 
ELAIS-S1  &  2630 & 931 & 324 & 234 & 44.0\% & -0.028 & -0.034 & 5.5\% & 6.5\%  & 17.6\% & 22.9\%\\
W-CDF-S & 4053 & 1393 &  435 & 312 & 30.0\% &-0.026 & -0.109 & 5.8\% & 7.5\% & 16.1\% & 23.9\%\\
LSS & 5242 &  1299 & 630 & 431 & 65.7\% & -0.130 & -0.183 & 4.8\% & 8.1\% &21.3\% & 28.7\%\\

\end{tabular}
\tablefoot{From left to right, the columns list: Sample field; the original number of X-ray sources; the number of sources in each field that were detected in the LS10 survey; have spectroscopic redshift $z > 0.002$; and were not already present in the training sample of {\sc PICZL} (i.e. previously unseen); the fraction of failed photo-z in XMM-SERVS z$_{\mathrm{phot}}$ = -99; and the bias $\langle \Delta z \rangle$, variance $\sigma_{\rm NMAD}$ and outlier fraction $\eta$ obtained with XMM-SERVS (S, excluding z$_{\mathrm{phot}}$ = -99) and {\sc PICZL} (P) photo-z.}
\label{tab:servs}
\end{table*}

The XMM-Spitzer Extragalactic Representative Volume Survey \citep[XMM-SERVS,][]{Mauduit2012} encompasses three key fields: the XMM-Large Scale Structure \citep[LSS,][]{Chen2018}, spannning 5.3 deg\textsuperscript{2} with a flux limit of $6.5 \times 10^{-15}$ erg cm$^{-2}$ s$^{-1}$ over 90\% of the survey area in the 0.5–10keV band \citep{savić2023}; the Wide Chandra Deep Field-South \citep[W-CDF-S,][]{Ni2021} and the European Large-Area ISO Survey-South 1 \citep[ELAIS-S1,][]{Ni2021}, covering approximately 4.6 deg\textsuperscript{2} and 3.2 deg\textsuperscript{2} \citep{brandt2018,scolnic2018}, limited to $1.3 \times 10^{-14}$ erg cm$^{-2}$ s$^{-1}$, respectively, each selected for their exceptional multiwavelength coverage and strategic alignment with the DDFs.

\begin{figure}[h!]
\centering
\includegraphics[width=9cm]{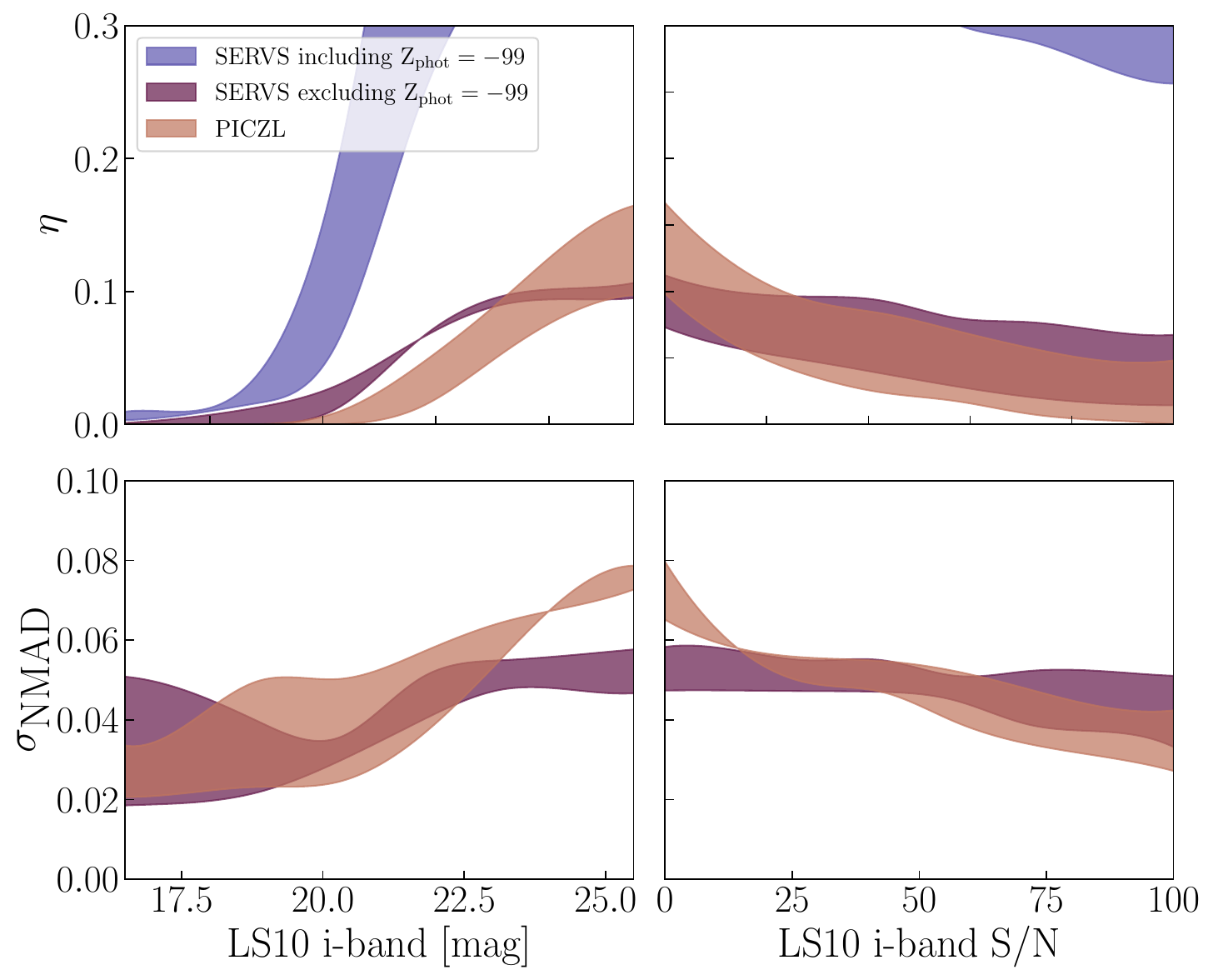}
\caption{Top: Outlier fraction as a function of LS10 i-band magnitude (left) and LS10 i-band S/N (right), comparing the original photo-z from XMM-SERVS to the photo-z computed with {\sc PICZL}. The XMM-SERVS photo-z results are shown both with and without the inclusion of sources for which XMM-SERVS photo-z failed (photo-z = -99). The shaded regions represent the range of XMM-SERVS outlier fractions across the ELAIS-S1, W-CDF-S, and LSS fields. Bottom: the range of XMM-SERVS variance in the three fields as a function of LS10 i-band magnitude and S/N. {\sc PICZL} demonstrates a lower outlier fraction for brighter sources while maintaining comparable accuracy to the original photo-z estimates. Additionally, {\sc PICZL} successfully computes photo-z for sources that are failures in the original methods.}
\label{fig:9.2}
\end{figure}

\subsection{Photo-z computation}

Considering that the original XMM-SERVS photo-z benefit from high S/N photometry and broader wavelength coverage, including $u$-band and NIR coverage, we would expect significantly better results compared to those achievable by {\sc PICZL} using LS10 data alone. However, {\sc PICZL} obtains comparable if not better results when limiting the sample to the depth at which it was trained on (see XMM-SERVS magnitude distribution in Figure \ref{fig:E}). Like any ML model, it struggles to extrapolate to faint sources outside its training distribution. Therefore, a fair comparison requires limiting the analysis to a similar feature space, while the results in Table \ref{tab:servs} reflect performance across the entire, diverse XMM-SERVS samples. To facilitate a comparison with {\sc PICZL}, our analysis begins with the counterpart associations of X-ray emissions as presented in \citet{Ni2021} and \citet{Chen2018}. Whenever possible, we limit our samples to sources flagged as AGN \citep[as defined in Sect. 6 of][]{Ni2021}. Matching to objects detected in LS10, due to its more limited depth, we identify approximately one-third of the original sources (see Table \ref{tab:servs}). This limitation is particularly pronounced in the XMM-LSS survey, where, in addition to deeper X-ray data corresponding to fainter counterparts, \cite{Chen2018} restrict their sample of AGN for which they calculate photo-z, to strictly non-broad-line X-ray sources. We select all sources with spectroscopic redshift and exclude those previously included in the {\sc PICZL} training sample (see Table \ref{tab:servs}). Unlike the original works, {\sc PICZL} also successfully computes photo-z for the significant fraction of sources for which the SED fitting adopted in XMM-SERVS failed (refer to w/photo-$z_{\mathrm{S}}$ in Table \ref{tab:servs}). Figure \ref{fig:9.2} visualizes the outlier fractions (top) and variances (bottom) for the aggregate XMM-SERVS samples. Blue and burgundy represent the distributions including/excluding catastrophic failures in the XMM-SERVS catalogs. The plot is limited to outlier fractions of 30\% and variances of 10\%, as XMM-SERVS curves including z$_{\mathrm{phot}}$ = -99 solutions continue to rise for fainter magnitudes or lower S/N. This suggests that, with increasing magnitude, a decreasing number of accurate estimates are available, making the subset excluding such solutions increasingly non-representative in terms of outlier fraction.

\begin{figure}[h!]
\centering
\includegraphics[width=9.1cm]{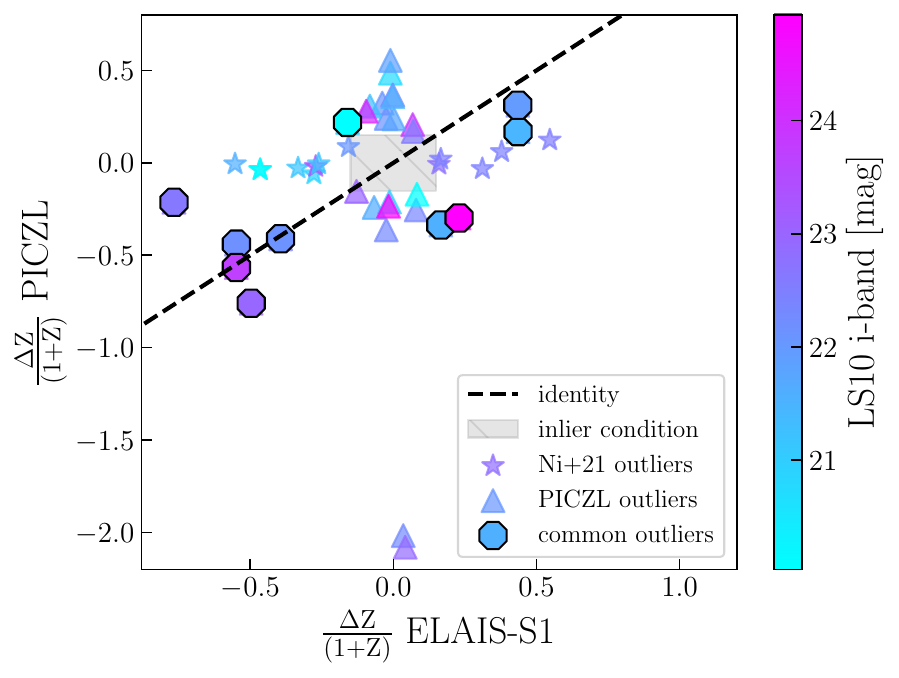}
\caption{Normalized residuals for {\sc{PICZL}} photo-z compared to XMM-SERVS photo-z, using data from XMM-SERVS ELAIS-S1. {\sc{PICZL}} outliers are shown as triangles, XMM-SERVS outliers as stars, and sources that are outliers in both methods are depicted as octagons. All sources are color-coded based on their LS10 i-band magnitude.}

\label{fig:9.1}
\end{figure}

However, by limiting our comparison to sources with photo-z obtained from XMM-SERVS, we find that the accuracy of these photo-z remains comparable up to magnitudes around 23.5 AB, despite the more limited data being used. Additionally, we observe enhanced performance in {\sc PICZL} photo-z for sources with S/N values $\gtrsim$ 60, depending on the specific XMM-SERVS sample. Importantly, sources with S/N $\gtrsim$ 10-20 mark a critical threshold range in LS10 where lower S/N values lead to an exponential rise in photometric errors (see Figure \ref{fig:C}). This trend is particularly evident for magnitudes fainter than 21.5 in the i-band \citep{saxena2024} and visible in all panels of Figure \ref{fig:9.2}.

\begin{figure*}[h!]
\centering
\includegraphics[width=18cm]{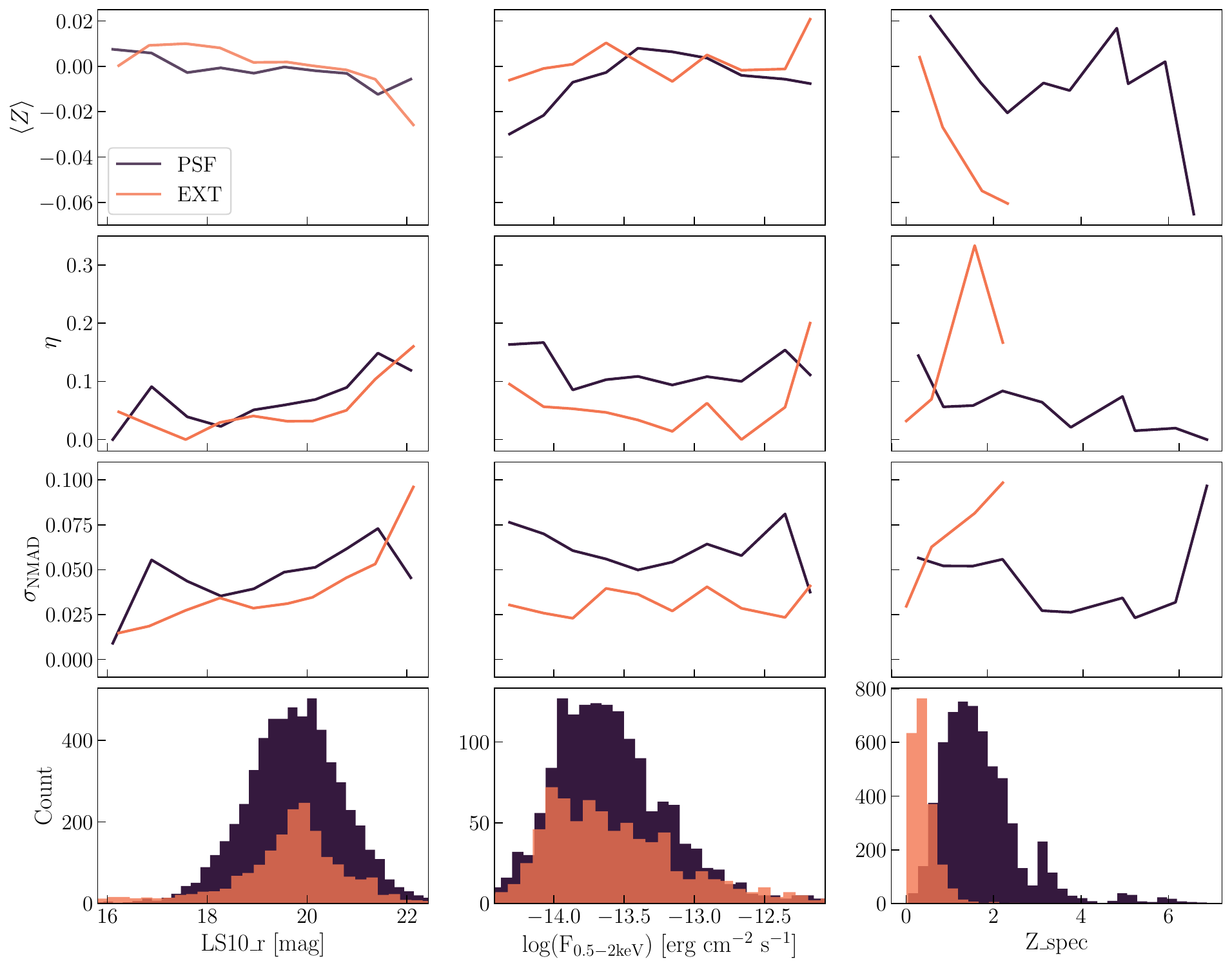}
\caption{Prediction bias (top row), fraction of outliers (second row), variance (third row), and number count (fourth row) as a function of magnitude ($r$-band from LS10), 0.2-2.3 keV X-ray flux, and spec-$z$, categorized by type (extended (EXT)/point-like (PSF)).}

\label{fig:4.5}
\end{figure*}

Figure \ref{fig:9.1} presents a visual comparison of normalized residuals for outliers identified by either {\sc{PICZL}} or the original photo-z from \cite{Ni2021}, exemplified via the ELAIS-S1 field. The spread of biases, excluding failures from XMM-SERVS, is similar for both methods, falling mostly within the range of [-0.5, 0.5]. The outliers in common appear to be modestly brighter than those that are outliers for a single method only and are evenly distributed between overestimation and underestimation. Notably, the outliers from \cite{Ni2021} are over a narrower range, while {\sc{PICZL}} has a few brighter and several more fainter, the majority of which with magnitudes beyond the range covered during training. The observed difference in outlier rates, is not primarily due to the use of template-fitting versus machine-learning methods but rather a reflection of the available photometry. XMM-SERVS, with its deeper photometry, is well-suited for faint objects, though saturation in brighter sources may contribute to some outliers. {\sc{PICZL}}, which leverages the LS10 dataset, is optimized for brighter sources due to the shallower photometry. While it performs well in this regime, fainter objects that fall outside the known parameter space are consequently less constrained.


\section{Discussion}
\label{sec10}

While \citet{saxena2024} made significant strides in overcoming the limitations of relying solely on total or model fluxes for photo-$z$ estimation of AGN by utilizing all redshift-correlated features in the LS10 catalog, we have further advanced this approach. By integrating data from both optical and MIR wavelengths, our study highlights the transformative potential of combining imaging with catalog-level data. Despite the state-of-the-art advancements in photo-$z$ estimation for AGN in this work, further improvements will be possible only by solving issues related to data quality. By overcoming these issues, we can significantly reduce the fraction of catastrophic outliers and achieve even greater accuracy in our predictions. In the following, we discuss the limitations that need to addressed in order to apply {\sc PICZL} to upcoming imaging surveys.

\subsection{Incorrect spec-z}
\label{ddi}

First and foremost, we need to address the reliability of the labels used in our, and generally any supervised machine learning algorithm. While spectroscopic redshift are typically trusted as accurate representations of true redshifts, this assumption does not always hold. As surveys continue to expand in scale, automated pipelines become indispensable for assigning spectroscopic redshift and identifying artifacts/problems for each source, as manual verification by visual inspection is impractical and will remain so for future surveys \citep{Bolton2012}. Importantly, most pipelines are trained on determining the redshift of the bulk of objects \citep{Alexander2023}, which are inactive galaxies and not AGN. This implies that a non-negligible fraction of AGN have wrong redshift estimates, including many labelled with "no warnings" \citep{Hewett2010,Wu2022}. Consequently, our performance is constrained by the unknown amount of incorrectly identified spectroscopic redshift per survey. While this fraction may be smaller for galaxies without nuclear activity, AGN present a range of challenges where pipelines fail. Examples of failures include cases where e.g. Mg-II is misidentified as Lyman-$\alpha$, leading to ambiguous redshifts around $z = 0.7-0.9 \Leftrightarrow 1.5-2.5$ or sources with strong featureless continuum typical of Blazars or, finally, sources with intense broad emission lines compared to the background continuum \citep{Chen2018,Ni2021}, as well as those with weak emission lines and a flat continuum background.

\subsection{Incorrect type classification}
\label{ddM}

\begin{figure}[t!]
\centering
\includegraphics[width=9cm]{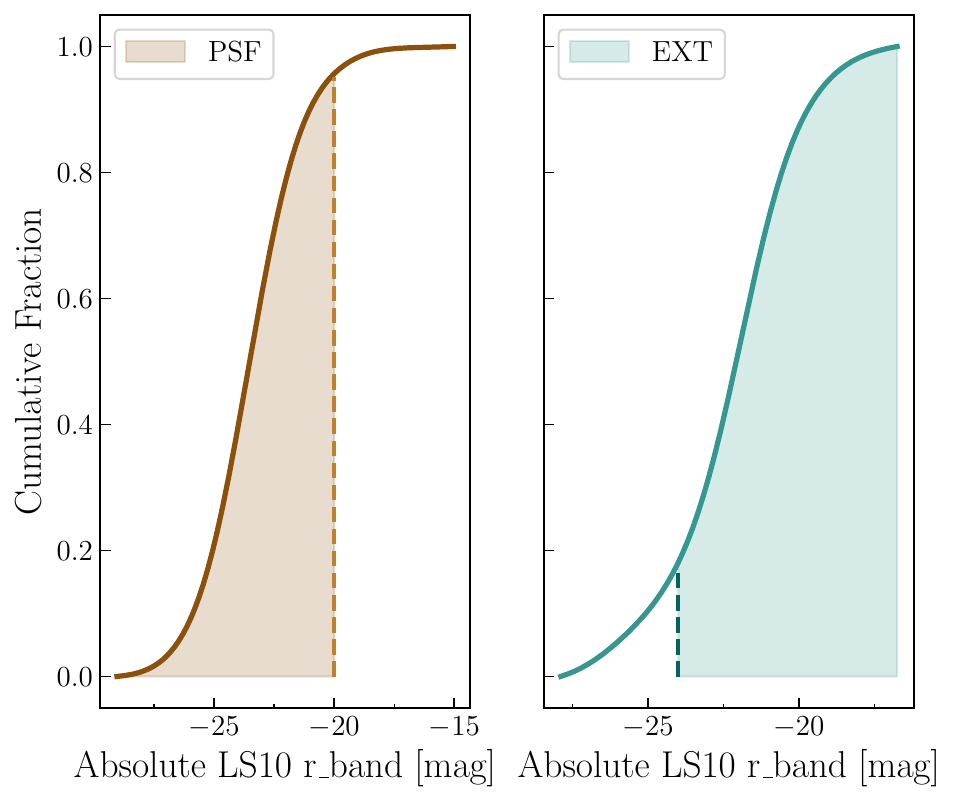}
\caption{Cumulative fraction of outliers as function of absolute magnitudes, to validate their classification based on TYPE. The left panel shows outliers classified as PSF, where most sources have absolute magnitudes consistent with AGN or QSO characteristics. In the right panel, 16\% of EXT-classified outliers exhibit absolute magnitudes more typical of QSOs.}

\label{fig:4.6}
\end{figure}

A subset of sources in the training sample, although having redshifts of $z \geq 1$, are classified as extended (i.e. non-PSF). This classification is likely incorrect, given the pixel resolution of LS10. The apparent extension is more plausibly a result of poor seeing conditions that were not properly accounted for, or the presence of nearby neighboring sources \citep[see Sect. 5.3 of][]{Hsu2014}. Since the morphological type affects predictions, akin to a prior in template fitting methods, this misclassification explains the increase in outliers within the EXT sample as redshift increases (see middle row in right panel of Figure \ref{fig:4.5}). To verify this, we computed the absolute {\it g} magnitudes for outliers to assess whether their values fall within the expected range for galaxies and AGN \citep[][]{Veron}. Figure \ref{fig:4.6} shows that 16\% of sources classified as EXT (and thus galaxy dominated) have an absolute magnitude which exceeds the range typical of galaxies M$_{\mathrm{EXT}}$ [-16,-24], confirming that their TYPE is wrongly assigned. Interestingly, approximately 96\% of PSF sources meet the absolute magnitude requirement of M$_{\mathrm{PSF}}$ [-20,-31], indicating that their TYPE is generally accurate.

\subsection{Faulty photometry}

In addition to issues connected to the reliability of labels, predictions also face challenges due to data inconsistencies arising from faulty photometry. Surveys often flag issues of this kind directly, including problems such as hot pixels, saturation, cosmic rays, bleed trails, and other uncategorized anomalies. Detecting and addressing such observational defects is crucial because the accuracy and reliability of photo-$z$ predictions cannot be guaranteed for compromised photometric inputs. We experimented with removing various groups of sources affected by observational artefacts but found no improvement when excluding any single group. We therefore assume that, while removing some problematic sources could potentially improve performance, the reduction in training data size offsets any gains. Adding this kind of noise to the current training sample could be an approach to decouple these two effects. Consequently, we treat the inclusion of all problematic sources as natural noise, as opposed to artificially degrading images employing methods such as Gaussian noise \citep{Hayat2021}.

\begin{table*}
\centering 
\caption{Comparison of bias, fraction of outliers, and variance between photo-z computed with \textsc{PICZL} for various subsamples.} 
\small
\begin{tabular}{clcccc}
\hline
\\
 Code & Sample & Number of sources &bias & variance & fraction of outliers \\
 & & & $\langle \Delta z \rangle $& $\sigma_{NMAD}$ & $\eta$ [\%]\\
 \hline
 \hline
 \\
{\sc{PICZL}} &validation   & 8098       &  - 0.004  & 0.045 & 5.6  \\ 
&    &        &   & &   \\

&Type: PSF  &  6083        & - 0.001  & 0.049 & 6.5  \\ 

&Type: EXT  & 2015    &   - 0.011 & 0.031 & 3.3  \\ 

&S/N > 3   & 6844         & - 0.002  & 0.044 & 5.4  \\ 

&S/N < 3   & 1254         & - 0.009  & 0.042 & 7.6  \\ 
&w/ i-band   & 7127         & - 0.002  & 0.044 & 5.6  \\ 
&w/o i-band   & 971         & - 0.011  & 0.044 & 7.1  \\ 
&optically selected AGN  & 4636  &  -0.002  & 0.040 & 3.9  \\ 
&X-ray selected AGN  & 3462  & - 0.005 & 0.050 & 8.9  \\
&isolated AGN   &    3934    &  0.005  & 0.042 &  5.5 \\
&w/ faint neighbour   &   407     &  0.004 & 0.051 & 6.4  \\
&w/ bright neighbour 
&   109     &  - 0.021 & 0.081 & 20.1  \\

&    &        &   & &   \\
& CSC2   &    416    &  -0.007  & 0.046& 8.3  \\
\\
\hline
\\
{\sc{CircleZ}}& CSC2   &   416    &  -0.009 & 0.055 & 12.3  \\

\\
\hline
\end{tabular}
\tablefoot{\tablefoottext{1}{all sources;}
\tablefoottext{2}{split by type (PSF vs. EXT);}
\tablefoottext{3}{split by signal-to-noise ratio (S/N) of all available bands;}
\tablefoottext{4}{split by availability of the {\it i}-band;}
\tablefoottext{5}{split by selection criteria;}
\tablefoottext{6}{split by the presence of a faint [(m$_{\mathrm{AGN}}$ - m$_{\mathrm{Neigh.}}$) within -3$\leq\Delta$mag<-1 for all available bands] or bright [(m$_{\mathrm{AGN}}$ - m$_{\mathrm{Neigh.}}$) within $\Delta$mag>-1 for all available bands] neighbor within a 5 arcsecond radius.} Additionally, results derived for the CSC2 blind sample for both {\sc{PICZL}} and {\sc{CircleZ}} are provided for comparison.}
 \label{tab:accuracyBreackdown}      
 \end{table*}

\subsection{Survey depth}
\label{Ph-errors}

When splitting the validation sample by S/N across all bands, we find a higher outlier fraction in sources with low S/N (see Table \ref{tab:accuracyBreackdown}). Consequently, to evaluate the reliability of our photo-z, we need to consider the photometric depth of LS10 and how increased photometric errors for fainter sources impact redshift estimate uncertainties. Figure \ref{fig:4.5} presents the prediction bias $\langle z \rangle$, outlier fraction $\eta$, and dispersion $\sigma_{\mathrm{NMAD}}$ as functions of $r$-band magnitude in LS10, X-ray flux (when available), and spec-$z$. As the $r$-band magnitude increases, as expected, the outlier fraction and scatter rise. This pattern is consistent with the exponential increase in photometric errors for, e.g. $i$-band magnitudes $\gtrsim$ 21.5, beyond which the outlier fraction, which otherwise remains well below 10\%, and scatter also appears to rise, especially for sources of type EXT (refer to Figure \ref{fig:C}). While faint LS10 sources have unreliable photo-z, future surveys such as LSST and Euclid will probe deeper with improved S/N. As such, our methodology can be adapted to these next-generation surveys, ensuring high accuracy across a broader range of magnitudes and providing robust redshift estimates for the extensive AGN populations these and other surveys will detect.

\subsection{Missing i-band}

For DR10, the Legacy Survey footprint was extended by incorporating all available DECam data from various contributing surveys (refer to Sect. \ref{sec3}), including coverage in the {\it i}-band, though limited to a single pass. As a result, $\sim$ 10\% of sources lack {\it i}-band observations. Unsurprisingly, and as shown in Table \ref{tab:accuracyBreackdown}, the accuracy of photo-z increases when this band is also available.

\subsection{Biases}

While the photo-z residuals exhibit a symmetric distribution around zero with minimal scatter, they do not consistently center at zero, suggesting a potential systematic bias in the photo-z estimates. Ideally, we would like to achieve normalized residuals that remain consistent irrespective of redshift or selection. Currently, this is not the case, as biases are not corrected for and are mostly influenced by the distribution of our training samples, which are specific to their respective survey and inherently biased.

\subsubsection{Selection effects}

We investigate whether the combination of various ways of selecting AGN entering the training sample affects the quality of the photo-z for specific subsamples. Using the classification outlined in Table \ref{Tab_2.1}, we split the validation sample based on observational criteria, creating a binary division between sources selected in optical or via X-ray. While there is only a minor difference in $\sigma_{\mathrm{NMAD}}$, we observe a higher outlier fraction for X-ray-detected sources. 

While strong X-ray emission from an AGN implies strong optical and mid-IR from an AGN, which should overpower the galaxy emission in the latter two bands, complicating the process of determining accurate photo-z, local Seyfert 1 galaxies with high X-ray flux have their host remain distinctly visible. As shown in the central panel of Figure \ref{fig:4.5}, however, the outlier fractions and accuracy remain consistent across the full range of X-ray fluxes, aside from small-number statistics for the wings of the distribution.

\subsubsection{Sample characteristics}

Unlike for galaxies, where the redshift distribution n($z$) is well-established and can serve as a reliable prior, the AGN n($z$) is not sufficiently characterised. Therefore uncertainty surrounds the ML algorithm subliminally adopting a distribution similar to the n($z$) of the training sample. 

At very low redshifts, AGN have a low surface density, resulting in only a few rare objects that can be considered for training. This phenomenon is intensified by the fact that deep surveys, as opposed to wide-field surveys, usually bypass such bright nearby objects in search of intermediate and high redshift sources. This leads to areas of scarcity in the training sample's spec-$z$ distribution, with biased predictions towards redshift values where more data points exist. To mitigate this, we normalize the CRPS score by (1+$z$), emphasizing the accuracy of low-redshift predictions. Additionally, at extremely low redshifts where the 4000 Å break is barely covered by the {\it g} filter,  accurate redshift predictions are more difficult to obtain. At very low redshift, the 6 arcsecond $\times$ 6 arcsecond cutout may be entirely filled by the galaxy, potentially misleading the algorithm into interpreting it as excess noise.

\subsubsection{Non-representative training samples}

One method to tackle covariate shift by the imbalance of the bright-end dominated spectroscopic sample, is given by subdividing both the training and validation samples into subsets of $n$-dimensional feature space of distinguishable properties \citep{NewmanGruen2022}. Specifically, the prediction accuracy improves if the model used to generate a posterior for a blind source was trained exclusively on training sources residing within the corresponding feature space \citep{Rosenbaum1984,Revsbech2017,autenrieth2023}. 
However, this approach might significantly reduce the training sample size per model making it more applicable to photo-$z$ codes dealing with large datasets, such as those for inactive galaxies \citep{NewmanGruen2022}.

In this work we have demonstrated that, in contrast to just a few years ago, AGN-targeted training samples are now sufficiently large to provide reliable photo-z. Overcoming inherent biases still requires either gathering more representative samples that uniformly cover the full redshift range—particularly underrepresented regions—or applying statistical corrections. Thus, the focus has shifted from simply increasing the number of spectra to strategically obtaining spectra that cover specific regions of parameter space \citep{Masters2015}. However, given the substantial gains our model has shown over existing methods,  a detailed examination of how these biases affect practical applications is beyond the scope of this study. Looking ahead, spectroscopic follow-up campaigns such as DESI, SDSS-V/BHM (Kollmeier et al., in prep), and 4MOST \citep{Jong2019} will further enhance our capabilities. Nevertheless, predicting photo-z for AGN in deeper Euclid and LSST datasets remains challenging due to the limited availability of spectroscopic data for faint sources. The upcoming Subaru Prime Focus Spectrograph \citep[PFS,][]{Tamura2016} and Multi Object Optical and Near-infrared Spectrograph for the Very Large Telescope \citep[VLT/MOONS;][]{Cirasuolo2020} are expected to help address this shortfall by providing much-needed spectra for faint (\& obscured/red) objects.

\subsection{Observational constraints: source crowding}

One factor beyond our control is the local environment or conditions along the line of sight, which can significantly influence the emission characteristics of AGN. Nearby objects can add extra flux, which can affect the accuracy of photometric measurements. We perform a positional cross-match between the validation sample and all LS10 sources within a 5 arcsecond radius of their optical counterparts. For each neighbour meeting this proximity criterion, we calculate apparent magnitudes and flag all sources where the brightest neighbour is no less than 1 magnitude dimmer. Table \ref{tab:accuracyBreackdown} shows that isolated sources exhibit better overall performance, while those with bright neighbours show drastically reduced quality compared to those with fainter neighbours. The presence of bright neighbours influences the observed flux in two ways: firstly, it complicates the derivation of source-specific colors due to convolved fluxes, and secondly, it affects the observed spectra, potentially leading to the detection of two sets of emission lines and incorrect spec-$z$ identification \citep{NewmanGruen2022}. A potential solution to this issue could be the implementation of segmentation maps, similar to SExtractor \citep[Source Extractor,][]{Bertin1996} at image level, as LS10 currently only masks neighbours during their model flux fitting procedure.

A persistent physical issue that cannot be entirely mitigated is blending. Unlike bright neighbours, which can be masked in principle, blending affects both observed photometry and spectroscopy. While particularly faint neighbours do not substantially affect the predictions negatively, sources identified as blends should be excluded from samples where accurate photo-$z$ estimates are required, until we can partially mask sources at the pixel level. The Rubin Observatory expects overlapping (inactive) galaxies to contribute at least 1\% of the total flux within their pixels \citep{Sanchez2021, NewmanGruen2022}. Blends are also expected to occur in the spectroscopic sample, increasing systematic uncertainty and subsequently the fraction of outliers by up to 5\% \citep{Masters2019}. This may be mitigated in future by higher-quality imaging from space-based data from missions such as Euclid or better tools for deblending \citep[e.g., {\sc SCARLET}/\texttt{blendz};][]{Melchior2018,Jones2018}. Alternatively, data augmentation or standard computer vision techniques could be implemented, such as artificially introducing blending effects in the training sample. However, this approach is not trivial, as LS10 currently lacks a corresponding blending flag.

\subsection{Variability}

AGN are inherently variable sources, meaning that their observed flux can change significantly across different epochs and wavelengths, especially when observations are separated by substantial time gaps. This extends to images that are created by stacking multiple observations taken over extended periods (as in LS10), where the resulting flux represents an average value. Such variability complicates the accurate prediction of AGN redshifts \citep{Simm2015}. However, future time-resolved imaging from LSST will allow us to better account for these variations and potentially use the correlation between AGN variability and their physical properties as a feature to improve photo-z predictions.

\subsection{Scalability in image-based photo-z}

Open issues remain regarding the feasibility of handling the vast data volumes and computational requirements associated with photo-z estimated from images. Exemplified by the calculation of new estimates for XMM-SERVS concerning storage, processing time, and computational resources, we give an overview in Appendix \ref{B}.

A crucial future direction involves exploring ways to utilize imaging data efficiently without necessitating extensive local downloads and computations for all image-based analyses of such data sets, potentially leveraging online platforms for real-time analysis, bringing the compute to the data \citep{Zhang2015}.


\section{Summary and outlook}
\label{sec11}

This work has been driven by the goal of determining reliable photo-z for X-ray detected AGN in wide-field surveys, such as eROSITA \citep{Merloni2024}. For that reason, we have concentrated on utilizing data from LS10, which provides sufficiently homogeneous coverage and depth in 4 optical bands, enriched by the flux measured on NEOWISE7 data for all identified sources (see Sect. \ref{sec3}). LS10 overlaps almost entirely with the eROSITA footprint, simplifying the cross calibration of data, a processing step typically necessary when merging different surveys. 

We introduce {\sc PICZL}, a CNN-based machine learning model designed primarily for AGN redshift estimation, but which holds the potential to reliably measure photo-$z$s for a broad range of extragalactic sources, provided an appropriately constructed training sample is available. In this study, the training sample comprises both X-ray and optically selected AGN. Across our validation set, {\sc{PICZL}} demonstrates consistently robust performance, with comparable accuracies and lower outlier fractions particularly for PSF sources, as opposed to the results obtained by \cite{Salvato22} for similar objects using SED-fitting (see Figure \ref{fig:4.4} and Table \ref{tab:accuracyBreackdown}). The results show significant improvements for both point-like and extended sources, indicating that the model's performance is primarily driven by the depth and hence photometric error of the training data. Additionally, when tested on a blind sample of X-ray-selected AGN, {\sc{PICZL}} maintained comparable results with respect to $\sigma_{\textrm{NMAD}}$ and $\eta$. Notably, on this test set it outperformed \citep[{\sc{Circlez}};][]{saxena2024} by achieving a 20\% improvement in accuracy and a 30\% reduction in the fraction of outliers (refer to Sect. \ref{sec6}).

We also applied {\sc{PICZL}} to estimate photo-z for the approximately 30\% of XMM-SERVS sources \citep{Chen2018, Ni2021}, detected in LS10 (see Table \ref{tab:servs}), demonstrating comparable variance and a substantially improved outlier fraction up to a limiting magnitude of 23.5 AB, using much fewer bands and significantly less sensitive imaging. However, it is important to note that the spectroscopic sample at this faint limit is small and likely biased toward sources with higher spectroscopic success rates. Consequently, the most reliable photo-z results are expected at brighter magnitudes \citep[as discussed in][]{saxena2024}. In response to these findings, we are releasing a new catalog of photo-z, complete with 1 and 3$\sigma$ error margins for the XMM-SERVS fields (see Sect. \ref{sec9}).

{\sc{PICZL}} will be used in the next generation of photo-z for sources detected by eROSITA, possibly switching from LS10 to Euclid and, most importantly, LSST, as they are poised to deliver more homogeneous and deeper data with broader wavelength coverage. In particular the availability of NIR data at image level, will improve our ability to determine accurate redshift for faint and high-z sources. Our study has shown that the performance of {\sc{PICZL}}, when relying predominantly on images, is already robust (see Table \ref{tab:loss}). Crucially, we have shown that well-calibrated images alone can suffice for accurate photo-z estimation, eliminating the need for catalog creation, which is often based on predefined models. This suggests that future surveys could reduce their dependence on catalog-based data for photo-z computation (refer to Sect. \ref{sec6}). Although this study has focused on calculating photo-z for AGN, the approach can be generalised and adapted to other source types, e.g., normal galaxies, with an appropriately constructed training sample (Götzenberger et al. in prep.). Subsequently our findings point to a bright future for all-sky surveys, also thanks to the continue expansion of training sample sizes, supported by initiatives like SDSS-V/BHM, 4MOST and VLT/MOONS, which will enhance the reliability and completeness of photo-$z$ estimates (refer to Sect. \ref{sec10}).

Looking forward, the next step in advancing photo-z estimation for AGN lies in exploring more sophisticated machine learning architectures beyond CNNs. For example, transformers with shared latent space embeddings offer a promising avenue. These models have shown success in integrating information from various data sources, such as images and entire spectra, potentially reducing uncertainties associated with traditional spec-$z$ pipelines by leveraging multi-modal data fusion \citep{DonosoOliva2023,Parker2024}. Additionally, incorporating other informative features like X-ray flux, when available, holds promise. Integrating these diverse data sources within a unified framework has the potential to refine redshift estimates even further.

\section{Data availability}

With this paper, we present a new catalog of photo-z derived using {\sc PICZL} for the $\sim$ 30\% of sources within the XMM-SERVS (ELAIS-S1, W-CDF-S, and LSS) X-ray source catalogs \citep{Chen2018,Ni2021} that are detected in the LS10 survey. Hence, it includes updated photo-z for sources with catastrophic failures in the original works. A detailed description of the catalog columns can be found in Appendix \ref{D}. The full catalog is only available in electronic form at the CDS via anonymous ftp to \url{cdsarc.u-strasbg.fr} (130.79.128.5) or via \url{http://cdsweb.u-strasbg.fr/cgi-bin/qcat?J/A+A/}.


\begin{acknowledgements}
      WR and MS are grateful for the constant support of Dustin Lang in handling Legacy Survey-related issues.
      Part of this work was supported by the German
      \emph{Deut\-sche For\-schungs\-ge\-mein\-schaft, DFG\/}, under Germany's Excellence Strategy – EXC 2094 – 390783311.
      We gratefully acknowledge funding from FONDECYT Regular - 1231718 (RJA), 1230345 (CR), and 1241005 (FEB), CATA-BASAL - FB210003 (RJA, CR, FEB), and ANID - Millennium Science Initiative - AIM23-0001 (FEB).
      JA acknowledges support from a UKRI Future Leaders Fellowship (grant code: MR/T020989/1)
       
This work is based on data from eROSITA, the soft X-ray instrument aboard SRG, a joint Russian-German science mission supported by the Russian Space Agency (Roskosmos), in the interests of the Russian Academy of Sciences represented by its Space Research Institute (IKI), and the Deutsches Zentrum für Luft- und Raumfahrt (DLR). The SRG spacecraft was built by Lavochkin Association (NPOL) and its subcontractors and is operated by NPOL with support from the Max Planck Institute for Extraterrestrial Physics (MPE).

The development and construction of the eROSITA X-ray instrument were led by MPE, with contributions from the Dr. Karl Remeis Observatory Bamberg \& ECAP (FAU Erlangen-Nuernberg), the University of Hamburg Observatory, the Leibniz Institute for Astrophysics Potsdam (AIP), and the Institute for Astronomy and Astrophysics of the University of T\"ubingen, with the support of DLR and the Max Planck Society. The Argelander Institute for Astronomy of the University of Bonn and the Ludwig Maximilians Universit\"at Munich also participated in the science preparation for eROSITA.

The Legacy Surveys consist of three individual and complementary projects: the Dark Energy Camera Legacy Survey (DECaLS; Proposal ID 2014B-0404; PIs: David Schlegel and Arjun Dey), the Beijing-Arizona Sky Survey (BASS; NOAO Prop. ID 2015A-0801; PIs: Zhou Xu and Xiaohui Fan), and the Mayall z-band Legacy Survey (MzLS; Prop. ID 2016A-0453; PI: Arjun Dey). DECaLS, BASS, and MzLS together include data obtained, respectively, at the Blanco telescope, Cerro Tololo Inter-American Observatory, NSF’s NOIRLab; the Bok telescope, Steward Observatory, University of Arizona; and the Mayall telescope, Kitt Peak National Observatory, NOIRLab. Pipeline processing and analyses of the data were supported by NOIRLab and the Lawrence Berkeley National Laboratory (LBNL). The Legacy Surveys project is honored to be permitted to conduct astronomical research on Iolkam Du’ag (Kitt Peak), a mountain with particular significance to the Tohono O’odham Nation.

NOIRLab is operated by the Association of Universities for Research in Astronomy (AURA) under a cooperative agreement with the National Science Foundation. LBNL is managed by the Regents of the University of California under contract to the U.S. Department of Energy.

This project used data obtained with the Dark Energy Camera (DECam), which was constructed by the Dark Energy Survey (DES) collaboration. Funding for the DES Projects has been provided by the U.S. Department of Energy, the U.S. National Science Foundation, the Ministry of Science and Education of Spain, the Science and Technology Facilities Council of the United Kingdom, the Higher Education Funding Council for England, the National Center for Supercomputing Applications at the University of Illinois at Urbana-Champaign, the Kavli Institute of Cosmological Physics at the University of Chicago, Center for Cosmology and Astro-Particle Physics at the Ohio State University, the Mitchell Institute for Fundamental Physics and Astronomy at Texas A \& M University, Financiadora de Estudos e Projetos, Fundacao Carlos Chagas Filho de Amparo, Financiadora de Estudos e Projetos, Fundacao Carlos Chagas Filho de Amparo a Pesquisa do Estado do Rio de Janeiro, Conselho Nacional de Desenvolvimento Cientifico e Tecnologico and the Ministerio da Ciencia, Tecnologia e Inovacao, the Deutsche Forschungsgemeinschaft and the Collaborating Institutions in the Dark Energy Survey. The Collaborating Institutions are Argonne National Laboratory, the University of California at Santa Cruz, the University of Cambridge, Centro de Investigaciones Energeticas, Medioambientales y Tecnologicas-Madrid, the University of Chicago, University College London, the DES-Brazil Consortium, the University of Edinburgh, the Eidgenossische Technische Hochschule (ETH) Zurich, Fermi National Accelerator Laboratory, the University of Illinois at Urbana-Champaign, the Institut de Ciencies de l’Espai (IEEC/CSIC), the Institut de Fisica d’Altes Energies, Lawrence Berkeley National Laboratory, the Ludwig Maximilians Universitat Munchen and the associated Excellence Cluster Universe, the University of Michigan, NSF’s NOIRLab, the University of Nottingham, the Ohio State University, the University of Pennsylvania, the University of Portsmouth, SLAC National Accelerator Laboratory, Stanford University, the University of Sussex, and Texas A\&M University.

BASS is a key project of the Telescope Access Program (TAP), which has been funded by the National Astronomical Observatories of China, the Chinese Academy of Sciences (the Strategic Priority Research Program “The Emergence of Cosmological Structures” Grant \# XDB09000000), and the Special Fund for Astronomy from the Ministry of Finance. The BASS is also supported by the External Cooperation Program of Chinese Academy of Sciences (Grant \# 114A11KYSB20160057), and Chinese National Natural Science Foundation (Grant \# 12120101003, \# 11433005).

The Legacy Survey team uses data products from the Near-Earth Object Wide-field Infrared Survey Explorer (NEOWISE), a project of the Jet Propulsion Laboratory/California Institute of Technology. NEOWISE is funded by the National Aeronautics and Space Administration.

The Legacy Surveys imaging of the DESI footprint is supported by the Director, Office of Science, Office of High Energy Physics of the U.S. Department of Energy under Contract No. DE-AC02-05CH1123, by the National Energy Research Scientific Computing Center, a DOE Office of Science User Facility under the same contract, and by the U.S. National Science Foundation, Division of Astronomical Sciences under Contract No. AST-0950945 to NOAO.

Funding for the Sloan Digital Sky Survey V has been provided by the Alfred P. Sloan Foundation, the Heising-Simons Foundation, the National Science Foundation, and the Participating Institutions. SDSS acknowledges support and resources from the Center for High-Performance Computing at the University of Utah. SDSS telescopes are located at Apache Point Observatory, funded by the Astrophysical Research Consortium and operated by New Mexico State University, and at Las Campanas Observatory, operated by the Carnegie Institution for Science. The SDSS web site is \url{www.sdss.org}.

SDSS is managed by the Astrophysical Research Consortium for the Participating Institutions of the SDSS Collaboration, including Caltech, The Carnegie Institution for Science, Chilean National Time Allocation Committee (CNTAC) ratified researchers, The Flatiron Institute, the Gotham Participation Group, Harvard University, Heidelberg University, The Johns Hopkins University, L’Ecole polytechnique f\'{e}d\'{e}rale de Lausanne (EPFL), Leibniz-Institut f\"ur Astrophysik Potsdam (AIP), Max-Planck-Institut f\"ur Astronomie (MPIA Heidelberg), Max-Planck-Institut f\"ur Extraterrestrische Physik (MPE), Nanjing University, National Astronomical Observatories of China (NAOC), New Mexico State University, The Ohio State University, Pennsylvania State University, Smithsonian Astrophysical Observatory, Space Telescope Science Institute (STScI), the Stellar Astrophysics Participation Group, Universidad Nacional Aut\'{o}noma de M\'exico, University of Arizona, University of Colorado Boulder, University of Illinois at Urbana-Champaign, University of Toronto, University of Utah, University of Virginia, Yale University, and Yunnan University.
\end{acknowledgements}

%
   \bibliographystyle{aa.bst} 
   \bibliography{bibliography}
%

\begin{appendix} 
\section{CNN model architecture}
\label{A}

\begin{figure}[h!]
\centering
\includegraphics[width=3.9cm]{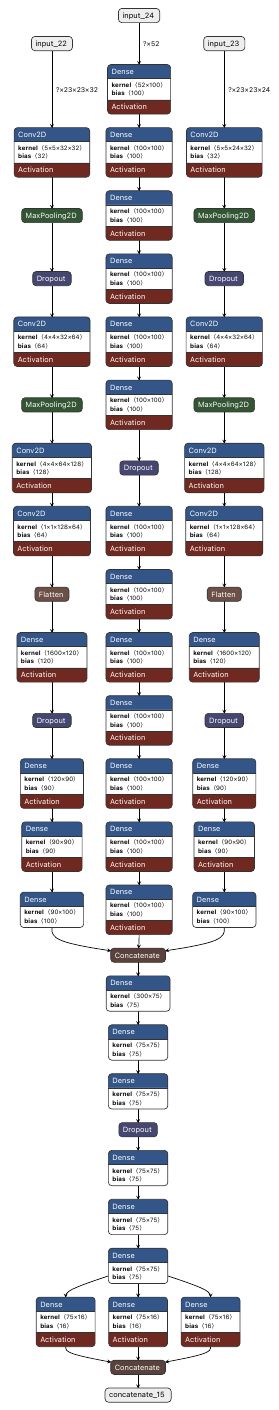}
\caption{{\sc {PICZ}L} architecture, showcasing its three-threaded design. Two threads are dedicated to processing image inputs, while the third handles numerical data. The model's output layer generates distinct vectors corresponding to the means, standard deviations, and weights of multiple Gaussian distributions. This design enables the production of full PDFs, allowing the network to capture uncertainties. Question marks as the first dimension per thread input correspond to the (variable) sample size.}

\label{fig:A.1}
\end{figure}

\section{LS10 seeing for different samples}
\begin{figure}[h!]
\centering
\includegraphics[width=9cm]{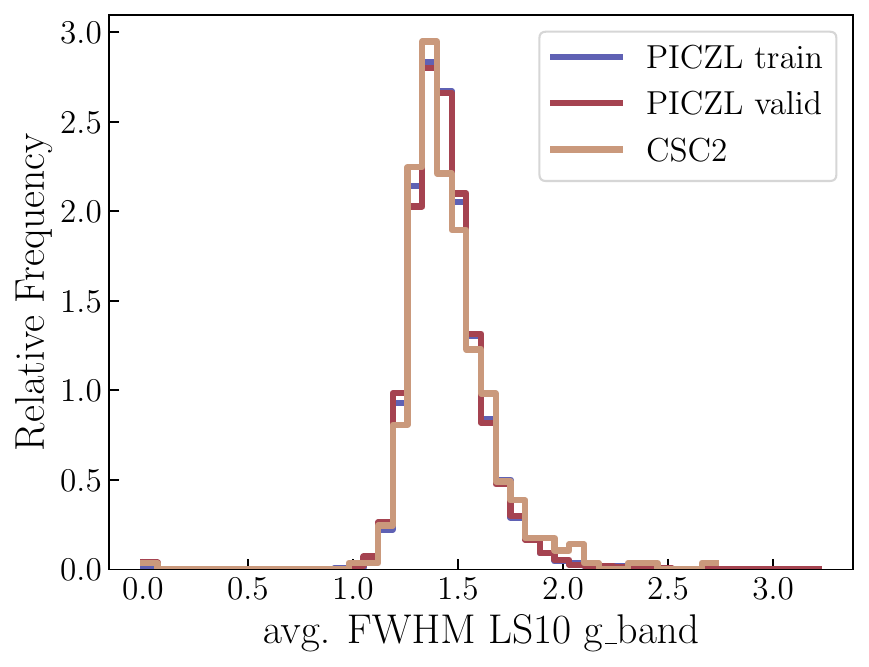}
\caption{Histograms of the weighted average FWHM (seeing) values for the {\sc{PICZL}} training, validation, and CSC2 blind samples. The close alignment of these distributions indicates no significant mismatch between the datasets, as they represent all-sky sampling. This similarity supports the applicability of the training sample to blind samples of comparable seeing.}

\label{fig:G.1}
\end{figure}

\section{CSC2 blind sample results}
\label{F}

\begin{figure}[h!]
\centering
\includegraphics[width=9cm]{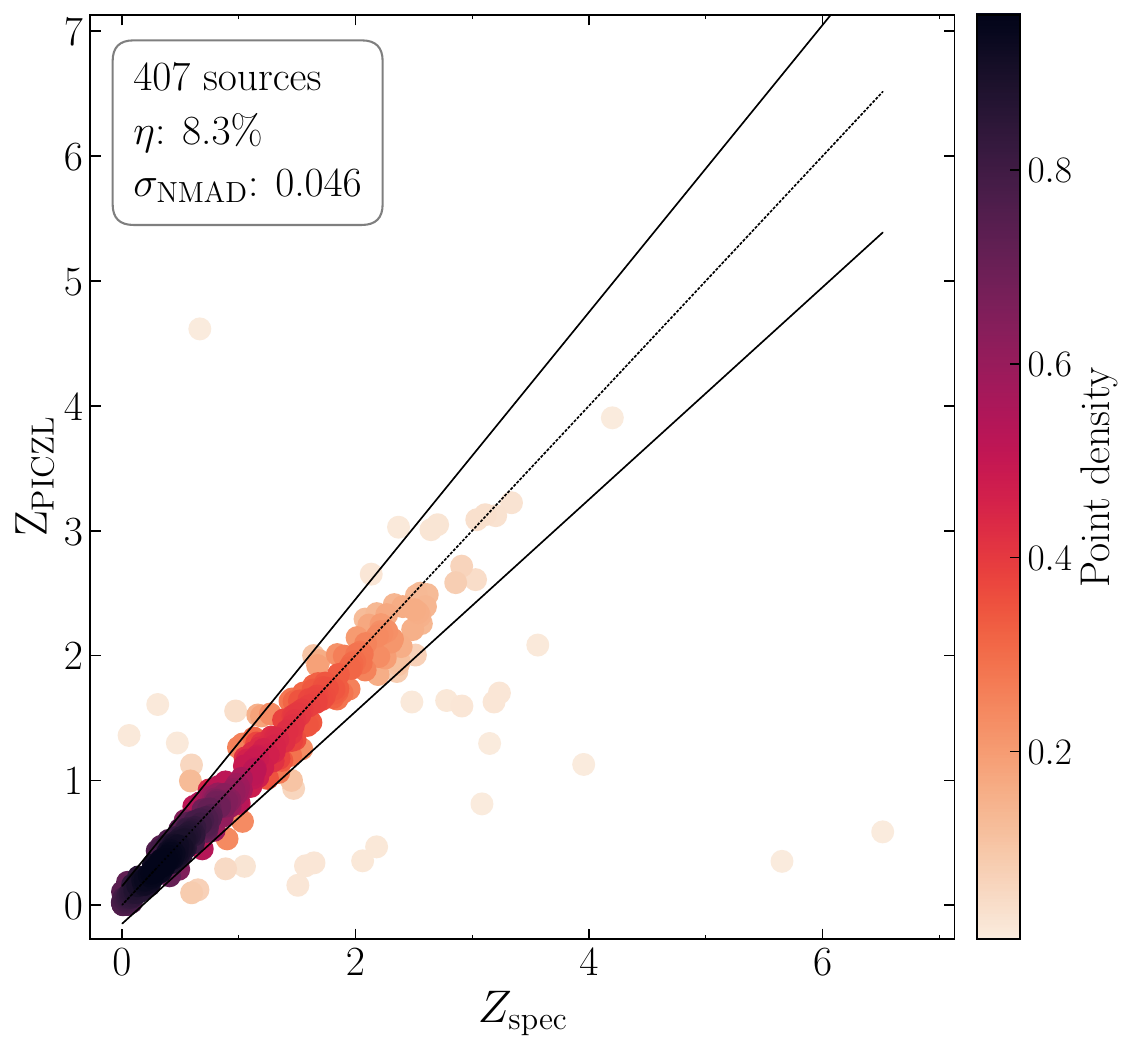}
\caption{Scatter plot of the CSC2 sample color-coded by point density, illustrating the performance of {\sc{PICZL}} on previously unseen data, as the majority of sources lie within the defined outlier boundary, demonstrating the prediction robustness. Sources identified as outliers generally scatter close to the boundary, with only a few instances of significant deviations or catastrophic failures.}

\label{fig:F}
\end{figure}

\section{Photometric errors in LS10}
\label{C}

\begin{figure}[h!]
\centering
\includegraphics[width=9cm]{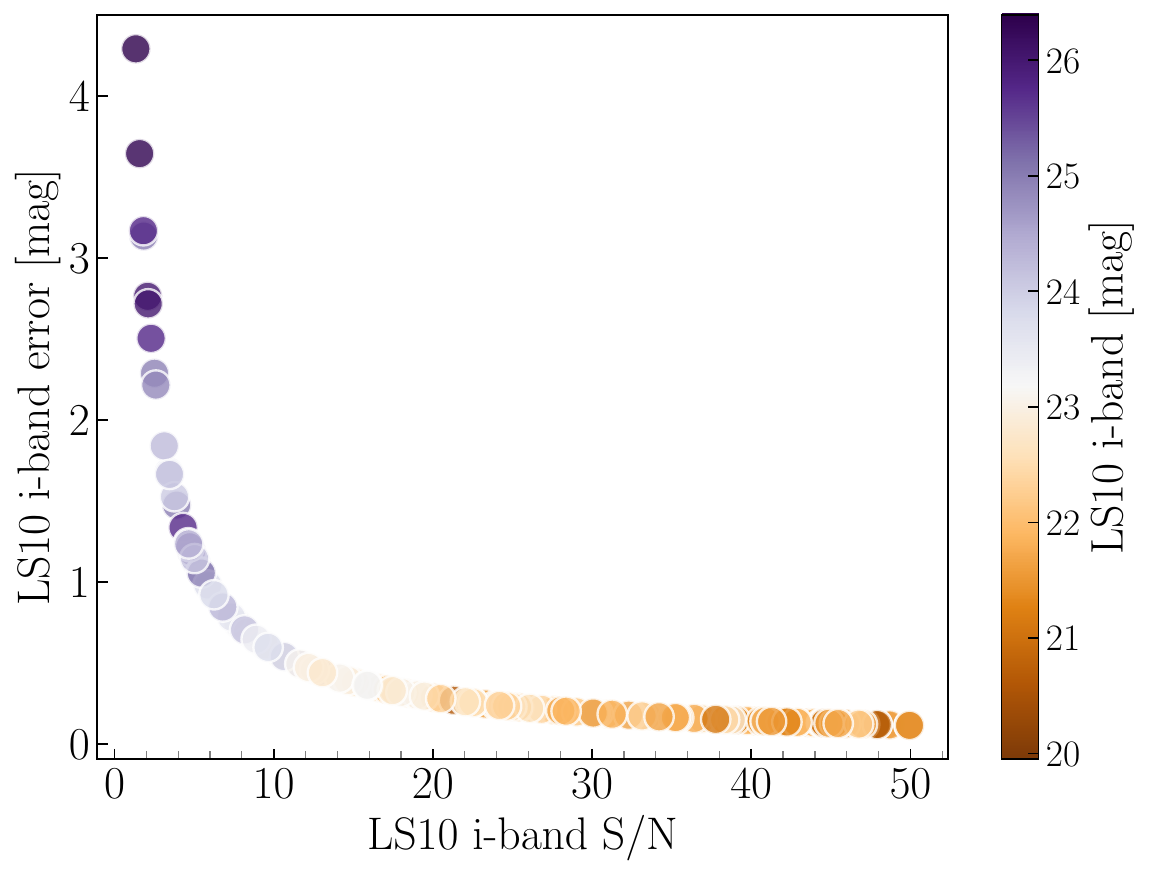}
\caption{Relationship between S/N and the associated photometric error for the LS10 $i$-band, considering all sources from Table \ref{Tab_2.1}. Each point is color-coded according to the $i$-band magnitude. A noticeable turning point is observed for sources with S/N $\lesssim 15$, indicating a significant increase in magnitude error for sources fainter than $\sim 22$ magnitude.}

\label{fig:C}
\end{figure}

\section{XMM-SERVS sources in LS10}
\label{E}

\begin{figure}[h!]
\centering
\includegraphics[width=9cm]{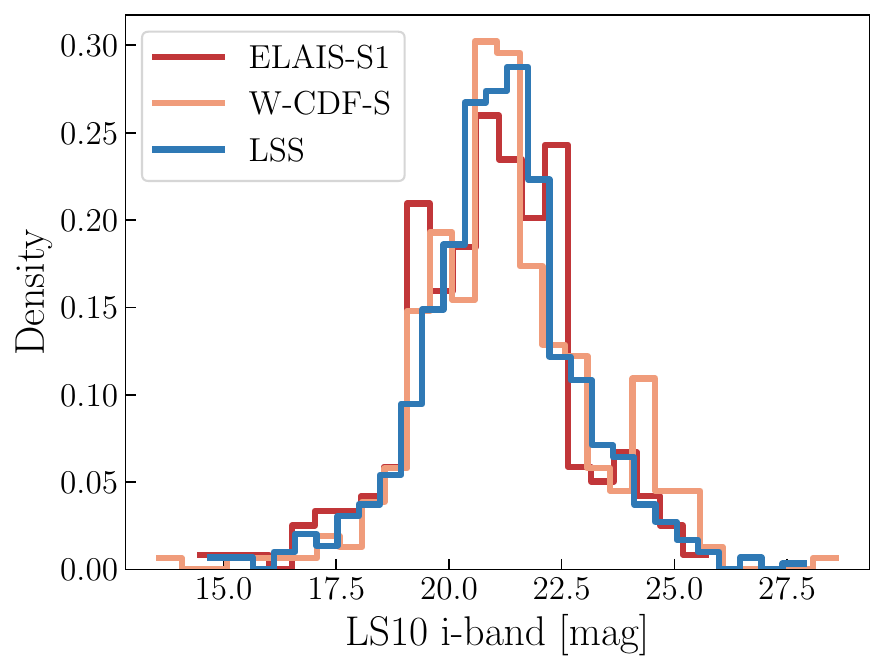}
\caption{Histograms of LS10 $i$-band magnitudes for the three XMM-SERVS subsamples: ELAIS-S1, W-CDF-S, and LSS. The magnitudes are displayed as normalized densities to allow for direct comparison across the fields. These distributions extend to fainter magnitudes than those shown in Figure \ref{fig:4.5}, accounting for the increased $\eta_{P}$ denoted in Table \ref{tab:servs}. As the depth reached in {\sc{PICZL}}'s LS10 training set is limited toward brighter magnitudes, a direct comparison between subsamples is only valid when conducted within a controlled parameter space, as shown in Figure \ref{fig:9.2}.}

\label{fig:E}
\end{figure}

\section{Column description of released photo-z catalog}
\label{D}

\begin{enumerate}
 \item XID: Unique source ID assigned to each X-ray source in the original papers \citep[see ][]{Chen2018, Ni2021}
  \item SURVEY: Original survey where the source was detected
  \item LS10\_FULLID: Unique LS source ID assigned to optical counterpart. It is created by concatenating the LS coloumns RELEASE, BRICKID and OBJID.
  \item CTP\_LS10\_RA: Right Ascension in degrees of the LS10 optical counterpart
  \item CTP\_LS10\_DEC: Declination in degrees of the LS10 optical counterpart
  \item SPECZ: spec-$z$ from original catalog
  \item PHZ\_PICZL: Photo-z from {\sc{PICZL}}
  \item PHZ\_PICZL\_l68: {\sc{PICZL}} photo-z min at 1 sigma
  \item PHZ\_PICZL\_u68: {\sc{PICZL}} photo-z max at 1 sigma
  \item PHZ\_PICZL\_l99: {\sc{PICZL}} photo-z min at 3 sigma
  \item PHZ\_PICZL\_u99: {\sc{PICZL}} photo-z max at 3 sigma
\end{enumerate}

\section{{\sc{PICZL}} run time}
\label{B}

The computational demands for training and deploying {\sc{PICZL}} depend mostly on the hardware configuration and dataset size. For our experiments, we leveraged a set of two Tesla V100-PCIE graphics processing units (GPUs) each equipped with 32 GB of ready access memory (RAM), accompanied by a 48 core multi-thread CPU to accelerate the computational workload. This parallel processing capability significantly expedited the model training process, compared to running solely on a central CPU, allowing for faster convergence. Training a single model (refer to Figure \ref{fig:A.1}) on a dataset of 32$\ $391 sources, corresponding to the 80:20 train-test split (refer to Table \ref{Tab_2.1}), each with 108 features (56 of which are images), takes approximately 25 minutes for 600 epochs. The use of an ensemble further scales the processing time by the number of models trained, with ensemble optimization depending on user preferences, making it difficult to provide a fixed time estimate. After finalizing the model, computing photo-z for e.g. 1393 sources in the XMM-SERVS W-CDF-S field, as displayed via Table \ref{tab:servs}, takes roughly 17 seconds. The storage requirements for the data used in this sample corresponds to approximately 250 MB (refer to Table \ref{Tab_2.2}). Looking ahead, upcoming surveys such as Euclid and LSST will produce higher-resolution images, demanding increased storage and computational resources due to larger data volumes and pixel counts. If the input dimensions were to increase, for instance, from 23x23 pixels to 64x64 pixels, the computational burden would rise significantly. While our current architecture of 32 GB RAM per GPU is efficient for the current input sizes, processing larger cutouts may necessitate either more GPUs or GPUs with higher memory capacity to maintain feasible training times and performance. In scenarios where larger input sizes are anticipated, a thoughtful approach to model architecture and resource allocation will be crucial. Therefore, adapting to the capabilities of the hardware will be key to successfully utilizing the methods in the context of future data from LSST and Euclid, particularly given the much larger sample sizes that we expect from these surveys. Finally, performance will also improve by having {\sc{PICZL}} revised by an expert software developer.


\end{appendix}

\end{document}